\documentclass[a4paper,fleqn,usenatbib]{mnras}

\usepackage{graphicx}
\usepackage{amsmath,amssymb}
\usepackage{enumitem}
\usepackage{placeins}
\usepackage[caption=false]{subfig}
\usepackage{longtable}
\usepackage{multirow}
\usepackage{booktabs}
\usepackage{color}
\usepackage{txfonts}

\def\Teff{$T_{\mathrm{eff}}$}
\def\logg{$\log\,g$}
\def\Mv{$M_{V}$}
\def\loggf{$\log\,gf$}

\def\Vt{$V_{\rm t}$}

\def\dech30{{\sc DECH30}}

\def\FeH{$\mathrm{[Fe/H]}$}
\def\Fe1{Fe\,{\sc i}}
\def\Mg1{Mg\,{\sc i}}
\def\Sc1{Sc\,{\sc i}}
\def\V1{V\,{\sc i}}
\def\Na1{Na\,{\sc i}}
\def\Cr1{Cr\,{\sc i}}
\def\Ni1{Ni\,{\sc i}}
\def\Ti1{Ti\,{\sc i}}
\def\S1{S\,{\sc i}}
\def\Na1{Na\,{\sc i}}
\def\Y1{Y\,{\sc i}}
\def\Mn1{Mn\,{\sc i}}
\def\Si1{Si\,{\sc i}}
\def\Co1{Co\,{\sc i}}
\def\C1{C\,{\sc i}}

\def\loggf{$\log\,gf$}

\title[Temperatures of classical Cepheids from LDR]
{Effective temperatures of classical Cepheids from line-depth ratios in the $H$-band}

\author[V.~Kovtyukh et al.]
{V.~Kovtyukh$^{1,2}$\thanks{This research used the facilities of the Italian Center for
Astronomical Archive (IA2) operated by INAF at the Astronomical
Observatory of Trieste, programme \mbox{OPT19A5}. This research is based [in part] on data collected at the Subaru Telescope, which is operated by 
the National Astronomical Observatory of Japan. We are honored and grateful for the opportunity of observing the Universe from Maunakea, which has the cultural, historical, and natural significance in Hawaii.
\newline E-mail: vkovtyukh@ukr.net},
B.~Lemasle$^{3}$,
N.~Nardetto$^{4}$,
G.~Bono$^{5,6}$,
R.~da Silva$^{6,7}$,
N.~Matsunaga$^{8}$,
A.~Yushchenko$^{9,1}$,
\newauthor
K.~Fukue$^{10,11}$,
and E.~K.~Grebel$^{12}$
\\
$^{1}$ Astronomical Observatory, Odessa National
University, Shevchenko Park, UA-65014 Odessa, Ukraine\\
$^{2}$ Institut f\"{u}r Astronomie und Astrophysik, Kepler Center for 
Astro and
Particle Physics, Universit\"{a}t T\"{u}bingen,\\
Sand 1, 72076 T\"{u}bingen, Germany\\
$^{3}$ Agentur f\"{u}r Arbeit Heidelberg
69108 Heidelberg, Germany  \\
$^{4}$ Universit\'e C\^ote d'Azur, Observatoire de la C\^ote d'Azur, CNRS, Laboratoire Lagrange, France  \\
$^{5}$ Dipartimento di Fisica, Universit\'a di Roma Tor Vergata, via della Ricerca Scientifica 1, I-00133 Rome, Italy\\  
$^{6}$ INAF -- Osservatorio Astronomico di Roma, via Frascati 33, Monte Porzio Catone, I-00078 Rome, Italy \\  
$^{7}$   Agenzia Spaziale Italiana, Space Science Data Center, 
 via del Politecnico snc, I-00133 Rome, Italy \\ 
$^{8}$ Department of Astronomy, School of Science, The University of Tokyo, 7-3-1 Hongo, Bunkyo-ku, Tokyo 113-0033, Japan\\ 
$^{9}$ Astrocamp Contents Research Institute, Goyang 10329, Korea \\ 
$^{10}$ Laboratory of Infrared High-resolution Spectroscopy (LiH), Koyama Astronomical Observatory, Kyoto Sangyo University,\\ Motoyama, Kamigamo, Kita-ku, Kyoto 603-8555, Japan\\
$^{11}$ Education Center for Medicine and Nursing, Shiga University of Medical Science, Seta Tsukinowa-cho, Otsu, Shiga 520-2192, Japan\\
$^{12}$ Astronomisches Rechen-Institut, Zentrum f\"{u}r Astronomie der Universit\"{a}t Heidelberg, M\"{o}nchhofstr. 12-14, 69120  Heidelberg, Germany\\}

\date{Accepted XXX. Received YYY; in original form ZZZ}

\pubyear{2022}

\begin{document}
\label{firstpage}
\pagerange{\pageref{firstpage}--\pageref{lastpage}}
\maketitle

\begin{abstract}
\par The technique of line depth ratios (LDR) is one of the methods to determine the effective temperature of a star. They are crucial in the spectroscopic studies  of variable stars like Cepheids since no simultaneous photometry is usually available. A good number of LDR-temperature relations are already available in the optical domain, here we want to expand the number of relations available in the near-infrared in order to fully exploit the capabilities of current and upcoming near-infrared spectrographs.
\par We used 115 simultaneous spectroscopic observations in the optical and the near-infrared for six Cepheids and optical line depth ratios to find new pairs of lines sensitive to temperature and to calibrate LDR-temperature relations in the near-infrared spectral range. We have derived 87 temperature calibrations valid in the [4\,800--6\,500]\,K range of temperatures. The typical uncertainty for a given relation is 60-70\,K, and combining many of them provides a final precision within 30-50\,K. We found a discrepancy between temperatures derived from optical or near-infrared LDR for pulsations phases close to $\phi$$\approx$0.0 and we discuss the possible causes for these differences. 
\par Line depth ratios in the near-infrared will allow us to spectroscopically investigate highly reddened Cepheids in the Galactic centre or in the far side of the disk.
\end{abstract}   


\section{Introduction}

\par The effective temperature \Teff is a fundamental parameter of stellar atmospheres. Therefore, deriving the effective temperature of a star is the most important step in the analysis of a stellar spectrum, which enables the determination of the chemical composition of the star and of its evolutionary status. 
\par Since by definition, the effective temperature is the temperature of a black body that produces the same total power per unit area as the observed star, \Teff\ can be derived directly by knowing the stellar luminosity and radius \citep[e.g.,][]{Davis1974}, making interferometric techniques the best tool at our disposal. Unfortunately, interferometric measurements are currently limited to nearby stars that do not cover yet the entire parameter space. Alternatively, the infrared flux method \citep[IRFM,][]{Blackwell1977} also provides \Teff\ and the angular radius of the star by combining its integrated flux and the infrared flux in a given band. Also, the Surface Brightness color relation in used to derive the angular diameter 
variation and the distance of Cepheids \citep{Nardetto2023}.
\par It becomes then possible to calibrate (spectro-)photometric techniques, for instance measuring the Paschen continuum (3647-8206\,\AA) to determine \Teff\ from stellar fluxes. Another (robust) method relies on \Teff-color calibrations \citep[e.g.,][]{Alonso1996,Bessell1998}. Photometric techniques to derive \Teff\ are however sensitive to the other atmospheric parameters of the star (for instance its metallicity [Fe/H] or its surface gravity \logg). Moreover, it is always difficult to obtain an accurate determination of the interstellar reddening, especially for faint, distant objects in highly extincted regions, for which we have started to obtain high-resolution spectra, in particular in the near-infrared (NIR) spectral domain.\\

\par Purely spectroscopic methods might then be preferred. For instance, fitting the profile of Balmer lines \citep[e.g.,][]{Gehren1981} provides a good \Teff\ diagnostic (although only below $\sim$8\,000\,K) thanks to their low sensitivity on \logg. Metal line diagnostics enable us to determine simultaneously the atmospheric parameters \Teff, \logg, and [Fe/H] (and microturbulent velocity \Vt\ for 1D-analyses), either by the means of their curves of growth \citep[e.g.,][]{Cayrel1963} or by ensuring that abundances of various lines from the same element show no trend with their excitation potentials (to constrain \Teff) or with their equivalent width (to constrain \Vt). They can be applied even in the case of pulsating variable stars like classical Cepheids, see for instance \citet{Kovtyukh1999}. Such techniques require however accurate determinations of the atomic parameters of the line (e.g., their oscillator strengths and damping constants) and they are sensitive to departures from the Local Thermodynamical Equilibrium (LTE). Continuous progress in our knowledge of the physics of stellar atmospheres and increased computing power now allows us to directly compare an observed spectrum with grids of synthetic \citep[e.g.,][]{Recio-Blanco2006} or empirical \citep[e.g.,][]{Ness2015} spectra.\\

\par The line depth ratios (LDR) method, which is based on the ratio of the depths of two lines having different sensitivity to \Teff{ \citep{Gray1991,Gray1994} presents the advantage of being free from reddening effects and provides a high internal precision ($\approx$10\,K). In FGK stars, the depths of low-excitation lines of neutral atoms are highly responsive to \Teff, while those of high-excitation lines are relatively insensitive to \Teff  \citep{Gray2005}.
LDR calibrations are available for dwarf and giant stars \citep[e.g.,][]{Strassmeier2000,Caccin2002,Kovtyukh2003,Biazzo2004,Biazzo2006,Kovtyukh2006,Biazzo2007}. Combining a large number of calibrations improves the precision of the temperature determination significantly. The concept of LDR has recently been expanded to flux ratios (FR) by \citet{Hanke2018}, focussing on small wavelength domains rather than the core of absorption lines, and with exquisite absolute calibration. They have been adapted to the specifics of Cepheids by \citet{Lemasle2020}.

\par \citet{Kovtyukh2000,Kovtyukh2007,Proxauf2018}   (see also \citealt{Biazzo2004,Biazzo2006})  calibrated LDR for Cepheids in the optical domain. \citet{Vasilyev2017,Vasilyev2018} have confirmed the validity of the line depth ratios approach using 2D numerical models of Cepheid-like variable stars, where non-local, time-dependent convection is included from first principles. Line depth ratios of Cepheids have paved the way for studying the distribution of metals in the Milky Way thin disk \citep{Andrievsky2002a, Andrievsky2002b,Andrievsky2002c,Andrievsky2004,
daSilva2016,daSilva2022,Genovali2013,Genovali2014,Genovali2015,Kovtyukh2005b,Kovtyukh2022,Lemasle2007,Lemasle2008,Lemasle2013,Luck2003,Luck2006,Luck2011a,Luck2011b,Luck2018b,Martin2015,Pedicelli2010}. Cepheids in the Magellanic Clouds also allow us to investigate the distribution of metals in the young population of these galaxies \citep{Lemasle2017,Romaniello2022}. Moreover, since the Large Magellanic Cloud is used to calibrate period-luminosity (PL) relations, LDR play a crucial role in investigating the possible metallicity dependence of PL relations \citep{Romaniello2008}. Finally, LDR have also been applied to old ($>$10Gyr) type II Cepheids \citep{Lemasle2015,Kovtyukh2018a,Kovtyukh2018b}, opening a new path to investigate thick disk and halo stars.
\par Cepheids' LDR also allowed us to trace temperature variations over the pulsation cycle \citep{Luck2004,Luck2008,Kovtyukh2005a,Andrievsky2005}, to discover peculiar Cepheids with high lithium content, presumably crossing the instability strip for the first time \citep[e.g.,][]{Kovtyukh2005c,Kovtyukh2019}, and to investigate Cepheids pulsating in two modes simultaneously \citep{Kovtyukh2016,Lemasle2018}.\\

\par The LDR method proved to be effective when applied to optical spectra, but it is only high-resolution IR spectroscopy that makes it possible to access the most distant stars in the Galactic disk and thereby understand the structure and evolution of the Milky Way in its innermost region, where interstellar extinction presents a serious problem \citep{Matsunaga2017}. The primary objects of surveys in this region usually have high luminosity -- namely, giants and supergiants. Recently, \cite{Fukue2015} found 9 LDR-\Teff\ relations using spectra of 8 stars (mainly giants) in the $H$-band (14\,000-18\,000\,\AA) for \Teff\ ranging from 4\,000 to 5\,800\,K with uncertainties of $\sim$60\,K. Later, \cite{Jian2019} increased the number of calibrations to 11 and achieved a precision of 35\,K for the range 3\,700$<$\Teff$<$5\,000\,K. Recently \cite{Afsar2023} report five new LDR-\Teff\ relations found in the $H$-band region and 21 new relations in the $K$-band. \cite{Taniguchi2018} found 81 calibrations for the \Teff\ within 3\,700$<$\Teff$<$5\,400\,K, using spectra of 9 giants in the $Y$- and $J$-band, and \cite{Jian2020} investigated the correlation between those calibrations and \logg. Subsequently, \cite{Taniguchi2021} obtained new LDR pairs of \Fe1-\Fe1\ lines for red giants and supergiants with \Teff\ of 3\,500--5\,500\,K. For spectra in the $Y$- and $J$-band, \cite{Matsunaga2021} developed a method for simultaneously determining \Teff\ and \logg\ for FGK stars of all luminosity classes; in so doing, they used 13 calibrations to deduce \Teff\ and 9 calibrations to derive \logg. All those calibrations were originally obtained in the IR range for the $Y$-, $J$-, $H$- and $K$-bands; however, they were only valid for rather low temperatures, while classical Cepheids reach \Teff\ above 6\,000\,K.

\par In this paper we want to expand the number of relations available for Cepheid studies in the near-infrared range. Sect.~\ref{data} describes the near-infrared spectra we used to search for new pairs of lines well-suited as temperature indicators, as described in Sect.~\ref{pairs}. The new LDR-\Teff\ calibrations are then investigated in Sect.~\ref{LDR}. Sect.~\ref{summ} summarizes our results.

\section{Spectroscopic material}
\label{data}

\par A large number of high-resolution spectra of six well-known bright classical Cepheids (Table~\ref{Cep}) were obtained with GIANO \citep{Origlia2014}, a NIR cross-dispersed echelle spectrograph, operating at the 3.6m Telescopio Nazionale 
Galileo (TNG). It covers the wavelength range  9\,500--24\,500\,\AA\ and operates at a very high-resolving power (R$\approx$50\,000). Optical spectra were obtained in parallel with the High Accuracy Radial velocity Planet Searcher North spectrograph \citep[HARPS-N,][]{Cosentino2012}. HARPS-N covers a large fraction of the optical range ($\Delta\lambda$=3\,900--6\,900\,\AA) at very high resolving power (R$\approx$100\,000).
The observing log is given in Table~\ref{indTeffCal}. 
\par Five additional spectra (3 of them for the calibrating Cepheids) were obtained in the $H$-band with the Infrared Camera and Spectrograph (IRCS) at the Subaru 8.2m telescope with a resolving power of R$\approx$20\,000 (\citealt{Kobayashi2000}, see Table~\ref{SUBARU}). Since we have no means to derive a priori their \Teff\, as we do not have simultaneous optical spectra for those stars, they were only used for testing the newly obtained relations. 

\begin{table}
\small
\caption{Parameters of the calibrating classical Cepheids.}
\label{Cep}
\begin{tabular}{rrrrrcccc}
\hline
Cepheid   &       P         & $<$V$>$  & $<$H$>$& \FeH&  \Mv   \\
          &       day       & mag      & mag    &  dex  &   mag   \\
\hline                                                         
 $\delta$ Cep&     5.3662   &     3.950&  2.479 &   0.07& --3.23 \\                   
   X Cyg  &       16.3512   &     6.399&  3.947 &   0.09& --4.52 \\                            
   S Sge  &        8.3823   &     5.618&  3.845 &   0.08& --3.75 \\                                
   T Vul  &        4.4355   &     5.751&  4.237 & --0.05& --3.01 \\                            
   S Vul  &       68.6510   &     8.972&  4.806 &   0.09& --6.19 \\                 
  SV Vul  &       44.8942   &     7.230&  4.051 &   0.11& --5.70 \\        
\hline
\end{tabular}
\end{table}

\par The spectral analysis (setting the continuum position, measuring line depths and equivalent widths) was carried out using the DECH software package  \footnote{\textbf{http://www.gazinur.com/DECH-software.html}}. The absorption lines of Cepheids are usually fairly broad due to pressure and Doppler broadening together with a moderate rotation ($\omega\leq$10\,km/s), and their a Voigt profile can be approximated by a Gaussian. However, they may become strongly asymmetric at some phases \citep[e.g.,][]{Nardetto2006,Nardetto2008b}. For this reason, we did not fit the entire profile but measured the line depths R$_{\lambda}$ (that is, between the continuum and a parabola fit of the line core) as described by \cite{Gray1994}. Typical number of data points on which we performed the parabolic fit is 4-5.
\par The $H$-band spectra, in particular, are heavily contaminated by the absorption features caused by the Earth's atmosphere when observed from ground-based facilities. We did not perform a telluric correction, which consists in removing telluric features from the spectra. Instead, we used only wavelength ranges known to be practically free of telluric lines. We cannot exclude, however, that a few spectral lines are slightly contaminated by telluric lines.

\section{Searching for temperature-sensitive line pairs}
\label{pairs}

\par With the recent development of near-infrared spectrographs, it has become possible to extend the use of line depth ratios as \Teff{} indicators to this domain. \citet{Fukue2015} were the first to provide calibration relations, in the $H$-band (1.50--1.65\,$\mu$m). However, the paucity of low-excitation lines in this wavelength range, together with the strong molecular bands and numerous telluric lines, limited the number of useful LDR pairs to nine. Such a small number limits the precision in \Teff{} to $\approx$50\,K in the most favorable cases, while precisions of the order of 5--15\,K can be routinely achieved in the optical thanks to a large number of available LDRs \citep[e.g.,][]{Kovtyukh2007,Proxauf2018}. Later on, \citet{Taniguchi2018} extended this number to 81 covering the $Y$- and $J$-band.

\subsection{Searching for useful lines}

\par In this study, we adopted a new approach: first, we selected two spectra of classical Cepheids with temperatures of about 5000 and 6200\,K, representative of the range of temperatures reached by this class of stars. Line depths were then measured for all the spectra, regardless of whether the lines were blended or not, also including lines that are not reliably identified. Only lines that could be measured both in the stars with \Teff\ of 5000\,K and 6200\,K were kept, in order to ensure that the final relations will be applicable over a broad \Teff\ range.\\

\par For these lines, we then computed the ratios of their depths, R$_{6200}$/R$_{5000}$ and split them into three groups showing significant (1), moderate (2), or slight (3) variations with \Teff. Pairs most likely suitable for further testing were chosen from the first and third groups. We set as an additional condition that the distance between two lines composing a given pair should not exceed 300\,\AA. This algorithm yielded 1500 potentially useful line pairs. Finally, the selected lines were measured in all the spectra. The 1500 potential relations were visually inspected and fitted with polynomial relations. Ultimately, only the 87 best calibrations, accurate to within 150\,K, were retained. They are shown in the Appendix (Figures~\ref{FigA1}-\ref{FigA5}).\\

\par Examining the atomic parameters of the lines in the selected calibrations allowed us to draw the two following conclusions:
\begin{itemize}[nosep]
\item[-] Even lines with similar excitation potentials of the lower level (EPL) can show a good correlation with temperature. This unexpected conclusion can be explained as follows: if we consider two lines with close EPLs, but different oscillator strengths (\loggf), then at a given \Teff\, the weaker line may be located on the linear part of the curve of growth, while the stronger line lies on the horizontal part. Thus the ratio of the depths of these two lines will be sensitive to \Teff. An example for such a pair of lines is given in Fig.~\ref{EPL}. As a consequence, such a calibration can only be used for a limited  range of \Teff; it presents however the advantage of being independent of luminosity (or \logg). Indeed, lines with different EPLs respond differently to \logg\ variations.

\begin{figure} 
\resizebox{\hsize}{!}{\includegraphics{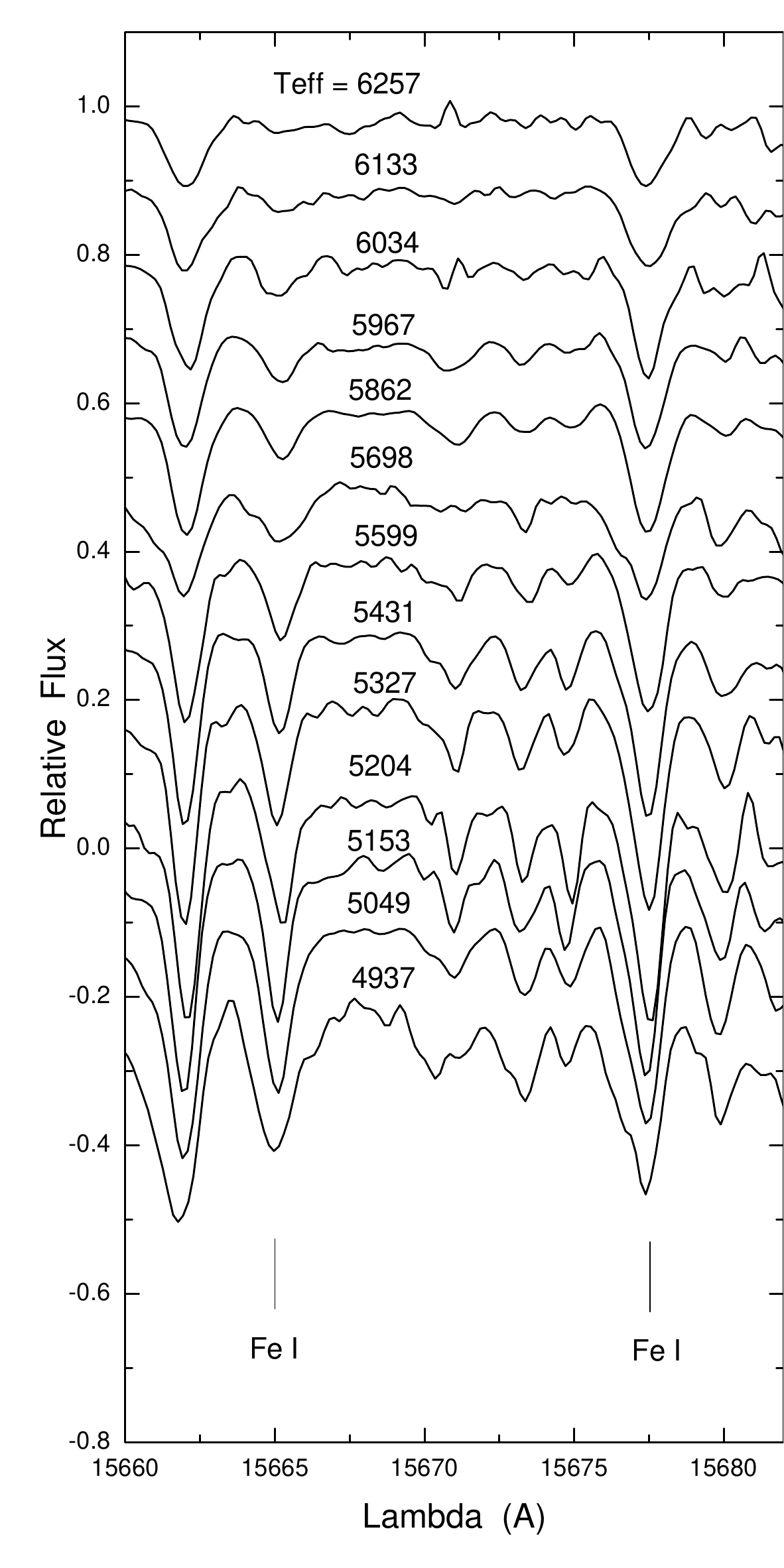}}
\caption{Variation of the line profile with \Teff\ for two lines with similar excitation potential of the lower level EPL, but with different oscillator strengths \loggf: the \Fe1\ line at 15665.240\,\AA\ (EPL=5.979\,eV, \loggf=--.336) and the \Fe1\ line at 15677.519\,\AA\ (EPL=6.246\,eV, \loggf=0.220). These two lines correspond to the calibration relation 80 (see Fig.~\ref{FigA5}).}
\label{EPL}
\end{figure}

\item[-] Although one would expect that only unblended lines should be considered (leaving only a small number of them available in the $H$-band, for instance), it is nevertheless possible to use strong blends to derive \Teff\ calibrations, provided that these blends change monotonically, gradually and unequivocally with the \Teff\ variations. An example for such a line pair is shown in Fig.~\ref{blends}, corresponding to the calibration relation 76 (see Fig.~\ref{FigA5})

\begin{figure} 
\resizebox{\hsize}{!}{\includegraphics{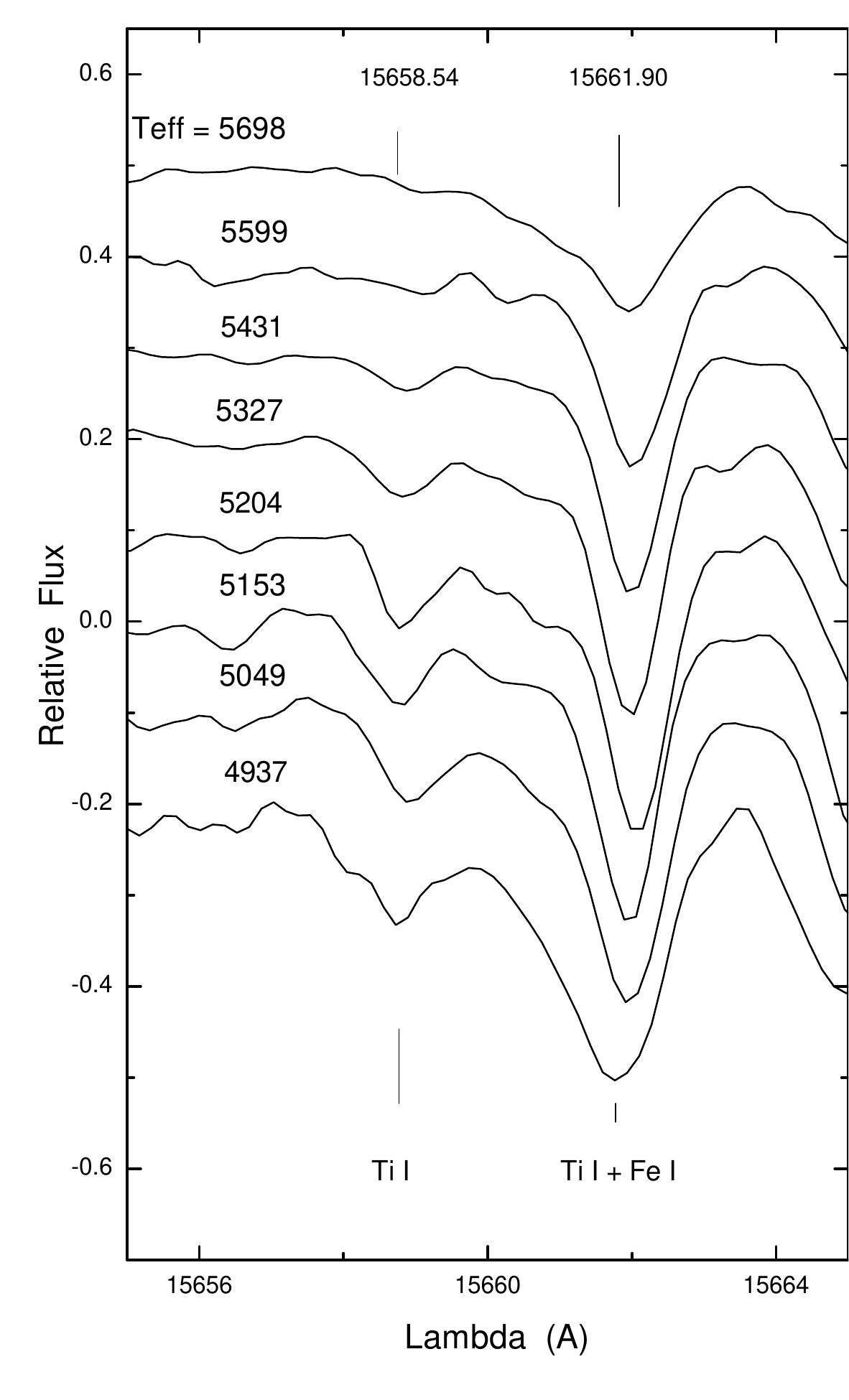}}
\caption{Variation of the line profile with \Teff\ for a calibration relation in which one of the lines forming the pair is blended: the first line of the ratio is the \Ti1\ line at 15658.545\,\AA\ (EPL=5.314\,eV, \loggf=--0.934), while the second line of the ratio is a blend of two lines at 15661.898\,\AA\ (\Ti1{}, EPL=5.172\,eV, \loggf=--0.550) and 15662.013\,\AA\ (\Fe1{}, EPL=5.828\,eV, \loggf=0.371). Together, they form the calibration relation 76 (see Fig.~\ref{FigA5}).}
\label{blends}
\end{figure}

\item[-] We note in passing that in case a telluric line would accidentally superimpose on a stellar line (which is more likely to happen in the $H$-band), the stellar line is discarded. Spectral lines of supergiants are usually considerably wider than the telluric lines, as shown in Fig.~\ref{tellurics}.  We did not use lines distorted by the influence of telluric lines. 

\begin{figure*} 
\resizebox{\hsize}{!}{\includegraphics{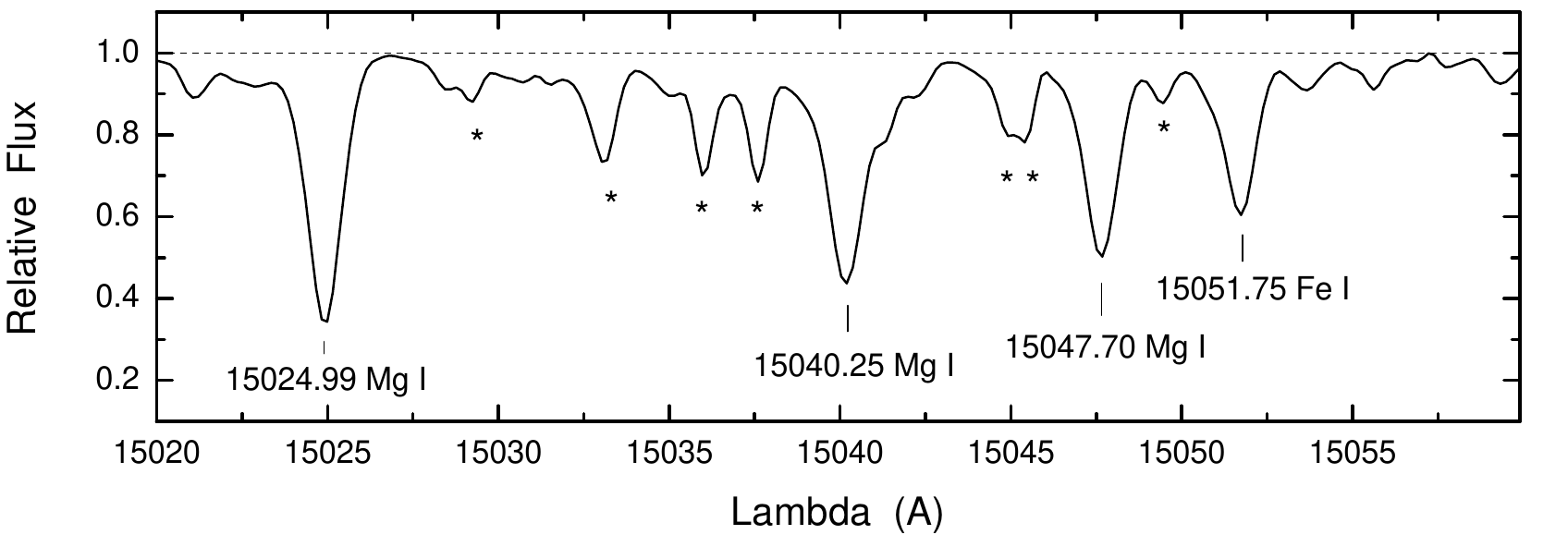}}
\caption{Fragment of a Cepheids' spectrum. The stellar lines are identified by their wavelengths and species, and the telluric lines are marked by asterisks.}
\label{tellurics}
\end{figure*}

\end{itemize}

\subsection{Calibrating relations}

\par For the TNG sample, the \Teff\ values used to calibrate the LDR relations have been derived from the optical HARPS-N spectra obtained quasi simultaneously. Indeed, the beginning of the exposures is shifted by only a few minutes, which is negligible since our 6 calibrating Cepheids have a period of $\approx$ 5 days and more. We used the LDR from \citet{Kovtyukh2007} (typically 50-60 of them are available in a given HARPS-N spectrum). This ensures that the new NIR LDR will fall on the exact same scale as those derived in the optical. We retained as calibrating \Teff\ the mean value of each temperature derived from a single calibrating relation in the optical, and the uncertainty on \Teff\ is the standard deviation of these measurements, usually around 10--30\,K.\\

\par With both line depth ratios and \Teff\ values at hand, it is possible to derive analytical formulae for new calibrating relations in the near-infrared. Polynomials offer a simple way to derive analytical relations, but a number of our calibrating relations show specific features such as breaks that cannot be adequately described even by polynomials of the 5$^{th}$ or higher degree (see Figs.~\ref{FigA1}--\ref{FigA5}). Therefore, we also tried more complicated relations, such as exponential fits, logarithmic fits, power fits, the Hoerl function ($y =a b^{x} x^{c}$) and others. The type of function (and the corresponding coefficients) that yielded the lowest root-mean-square deviation   $\sigma$  for a given calibration was ultimately selected. 

\par In many cases, the precision of an individual calibration relation varies with \Teff. This is related to the fact that the line strengths vary with temperature. For instance, at high \Teff, absorption lines with low EPL become weaker, leading to greater uncertainties in the measurement of their depths, until they eventually disappear from the spectrum. To take this effect into account, we have defined as an optimum range for a given ratio the \Teff\ range within which the  mean  precision   ($\sigma$)  of the calibration relation remains within 160\,K. Since various relations have various optimum ranges, we note that only a (large) subset of the 87 relations can be used for a Cepheid at a given temperature. This also holds for optical spectra and explains why the number of optical relations used to determine \Teff\ from the HARPS-N spectra varies from star to star.\\

\par Uncertainties on the line-depth measurements mainly arise from uncertainties in setting the continuum position, hence the presence of noise or telluric lines. It can be determined from
lines that fall twice on adjacent orders of the echelle spectra.
This uncertainty is about 2-6\% for spectra with a signal-to-noise ratio of about 100. 
  A complete analysis of the errors associated with measuring line depths in spectra is given in \cite{Catalano2002}. 
\par Besides, individual stellar parameters such as metallicity, rotation, convection, NLTE effects, magnetic fields, binarity, etc., add to the scatter of the individual calibrations. An analysis of such effects was presented in the studies by \citet[e.g.,][]{Gray1989,Gray1994,Strassmeier2000,Fukue2015}. 
\par The list of the calibrating relations, including the values for the coefficients, the intrinsic dispersion, and the applicability range, is given in Table~\ref{tableA1} (Appendix). They are displayed in Figs.~\ref{FigA1}-\ref{FigA5}.

\section{Testing the new line depth ratios in the NIR}
\label{LDR}

\par The temperatures inferred from both the optical and NIR spectra and their respective uncertainties are given in Table~\ref{indTeffCal}.
 A direct comparison of these temperatures is shown in Fig.~\ref{compar}. As can be seen, the agreement is excellent, and the largest deviations are localized for the highest \Teff\, above 6200\,K. 

\begin{figure} 
\resizebox{\hsize}{!}{\includegraphics{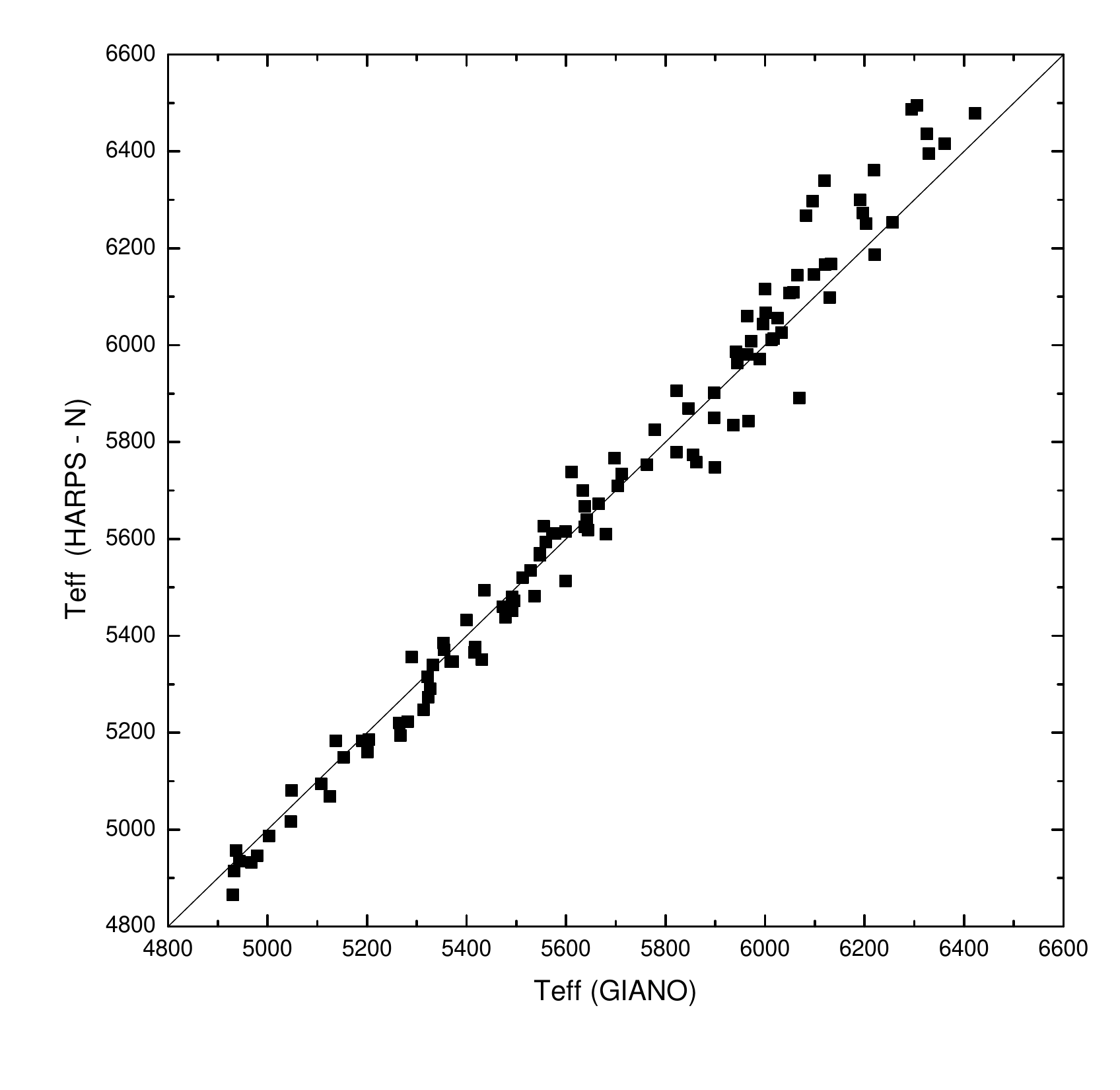}}
\caption{Comparison between the calibrating (optical, HARPS-N) and the retrieved (NIR, GIANO) temperatures. The 1:1 line is shown as a reference.}
\label{compar}
\end{figure}

\par Fig.~\ref{cepvar} provides an alternative look to the same data, displaying the variations of \Teff\ with the pulsation phase, where \Teff\ was computed with either the HARPS-N calibration data in the optical or the new NIR relations. The latter are provided for both the GIANO spectra and the SUBARU spectra, when available (see also Table~\ref{SUBARU}). Also with such a point of view, the agreement remains excellent, with the largest deviations being confined to phases close to $\phi$=0.0 where \Teff\ is maximal.\\

\begin{table}
\caption{Temperatures and associated uncertainties determined with the new LDR in the NIR for SUBARU spectra of Cepheids. The number of LDR used as well as the observing log is also provided.}
\label{SUBARU}
\begin{tabular}{lccccccc}
\hline\hline 
  Cep    &  P & JD       & phase & \Teff&$\sigma$&  N    &  $\sigma/\sqrt{N}$   \\
         & (d)&2\,400\,000+&       &   (K)  &  (K)&      & (K)   \\
\hline
CF Cas   & 4.8750   &  56137.110     & .859  &   6047 &  98 & 18   & 23.1    \\
DL Cas   & 8.0003   &  56137.080     & .461  &   5527 & 116 & 46   & 17.2    \\
$\delta$ Cep& 5.3662&  56136.114     & .260  &   5843 & 127 & 17   & 30.7   \\
X Cyg    & 16.3512  &  56136.105     & .975  &   6023 & 162 & 22   & 34.6    \\
SV Vul   & 44.8942  &  56137.058     & .231  &   5760 & 142 & 21   & 31.0   \\
\hline
\end{tabular}
\end{table}

\begin{figure*} 
\resizebox{\hsize}{!}{\includegraphics{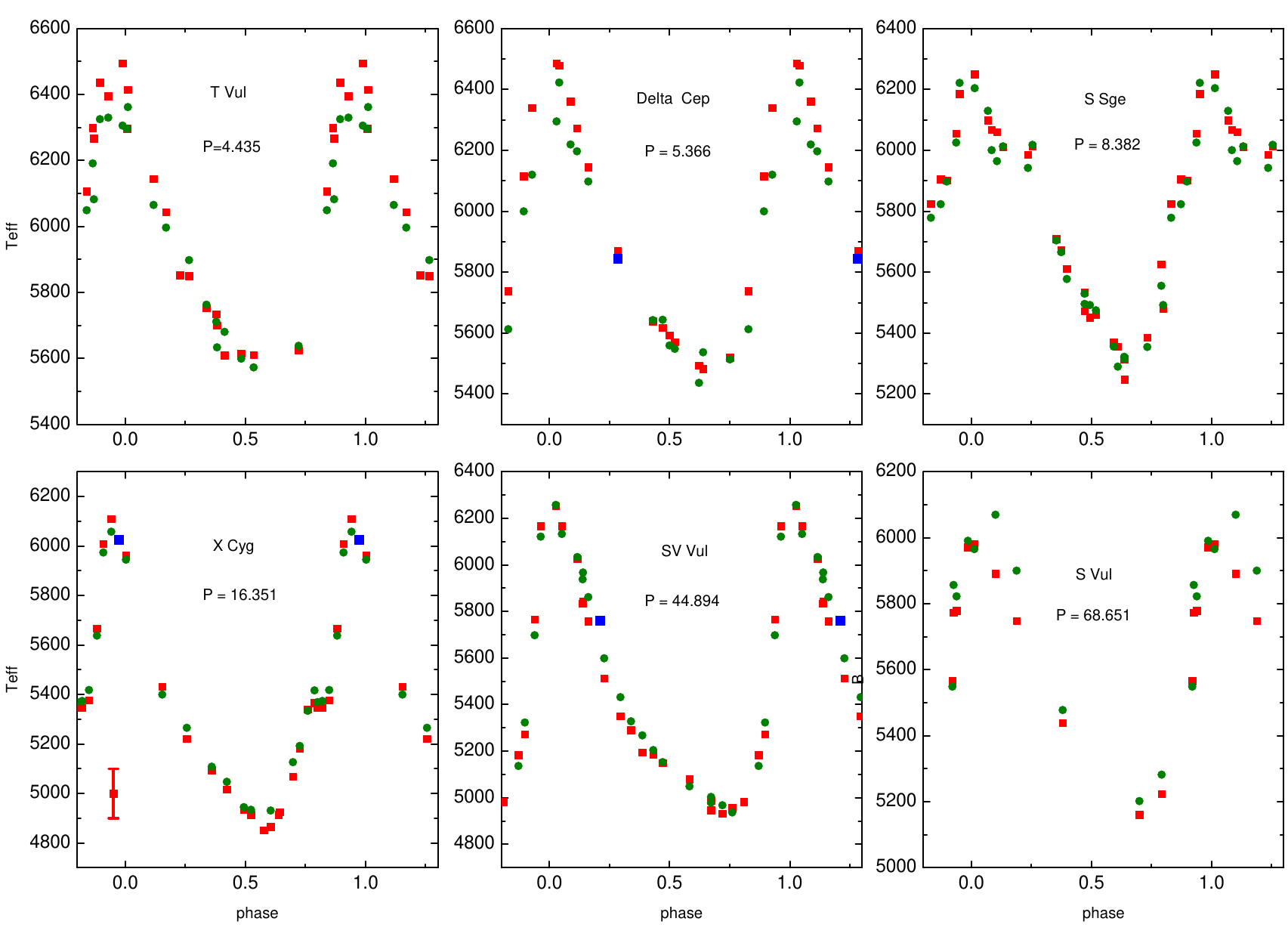}}
\caption{Temperature variations for the six Cepheids in the calibrating sample. Red squares: HARPS-N (optical) temperatures. A Fourier smoothing through the data points is indicated as a thin red line to guide the eye. Green circles: GIANO (near-infrared) temperatures. Blue squares: SUBARU (near-infrared) temperatures. A typical uncertainty  ($\sigma$)  is shown in the lower-left corner of the figure. The standard errors  $\sigma/\sqrt{N}$  on individual \Teff\ measurements are smaller than the symbol sizes.}
\label{cepvar}
\end{figure*}

\par We first notice that for T~Vul, a classical Cepheid with the shortest period in the calibrating sample (and hence, the lowest luminosity and the largest surface gravity), the NIR temperatures near the \Teff\ peak are systematically lower than those deduced from optical spectra. Conversely, for S~Vul, the long-period classical Cepheid with the highest luminosity (lowest surface gravity) in the calibrating sample, the NIR temperatures are higher than those deduced from optical LDR. This points toward a luminosity (or \logg) effect on the line depths ratios.  Several (related) explanations can be proposed for such behavior. 
\par \cite{Jian2020} already detected the effect of surface gravity on LDR. Indeed, for several pairs of lines, they noticed that the LDR--\Teff\ relations were offset between dwarfs on one hand, and giants and supergiants on the other hand. They found that the difference between the ionization potentials of lines in a given pair correlates with the sensitivity of this pair to \logg. A detailed theoretical analysis of this effect can be found in \citet{Gray2005} and \citet{Jian2020}. In order to circumvent this drawback, they suggested calibrating separately dwarfs and giants/supergiants.
\par However, in contrast with \citet{Jian2020}, who report no \logg\ effect within the giants-supergiants group, we find here significant \logg\ effects (for a narrow range of pulsation phases) for Cepheids, that is, for stars within the giants/supergiants luminosity class. We note however that the range of luminosities for the six Cepheids in our calibrating sample is very wide and amounts to three magnitudes (their absolute magnitudes vary from --3 to --6 \Mv,  see Table~\ref{Cep}). This may indicate that the effect in Cepheids is not, or not only, a \logg\ effect. For instance, \Teff\ values may also differ due to the differences in the optical depths of the line-forming regions for the optical and IR ranges. These differences can be significant at given pulsation phases, for instance, due to the shock wave passing through the upper atmospheric layers of Cepheids near the maximum compression. 
\par Indeed, \citet{Nardetto2018} investigated CRIRES observations of the long-period Cepheid $l$\,Car and found, using an hydrodynamical model of this star \citep{Nardetto2007}, that the core of the \Na1 line at 22\,089,69\,\AA\ is formed at the top of the atmosphere, while the iron lines in the visible are formed much deeper in the atmosphere. They report additional evidence that lines in the infrared are formed closer to the surface of the star than lines in the optical, for instance the infrared radial velocity curve is shifted with respect to its optical counterpart, which they interpret as a manifestation of the Van Hoof effect \citep{VanHoof1953}, the delay in the velocities between lines forming in the lower and upper atmosphere. Similarly, the mean radial velocity derived from infrared data differs by 0.53$\mp$0.30 km s$^{-1}$ from the optical one, which they interpret as a different impact of granulation on line forming regions in the upper and lower atmosphere \citep[the deeper the line-forming region, the more the radial velocity is blueshifted, see][]{Nardetto2008a,Vasilyev2017}.
\par To wrap things up, it seems established that visible and infrared lines are formed at different depths in the atmosphere of a Cepheid, and thus in environments in which not only temperature and pressure are different, but also the velocity fields (due to the propagation of the compression wave). The latter is clearly visible in the different behaviour of line asymmetries for optical and infrared lines over the pulsation period \citep[][ their Fig.\,6]{Nardetto2018}. Since we measure the line depths directly, without fitting a line profile, we assume that the differences between short- and long-period Cepheids at phases $\phi$$\approx$0.0 we observe in Fig\,\ref{cepvar} mostly reflect temperature differences rather than uncertainties on measuring line depths related to different line asymmetries.
\par Furthermore, we note that the theoretical analyses described in \citet{Gray2005} and \citet{Jian2020} are made under the Local Thermodynamical Equilibrium (LTE) assumption, while \citet{Vasilyev2018,Vasilyev2019} have shown that NLTE effects are important in the atmospheres of Cepheids and maximal at the same phases ($\phi$$\approx$0.0) where the discrepancy between optical and NIR line depth ratios is significant.
\par Finally, it is worth mentioning that long-period Cepheids are known to exhibit cycle-to-cycle variations \citep[e.g.,][]{Anderson2016}, including in their line profiles. However, this phenomenon cannot be invoked here since our optical and NIR have been observed simultaneously. Should long-period Cepheids be excluded from the calibration of the LDR, then their \Teff\ could not be determined and hence their chemical composition would remain unknown.

\section{Summary and Conclusion} 
\label{summ}

\par In the present study, we have derived 87 temperature calibrations, LDR-\Teff, using GIANO high-dispersion near-IR $H$-band spectra covering the wavelength range from 14\,000 to 16\,500\,\AA\ that contains numerous atomic lines and molecular bands. The temperatures inferred from the optical spectra obtained in parallel with the HARPS-N spectrograph were adopted as original temperatures to derive calibration relations. The resulting temperature relations are based on 115 spectra of six classical Cepheids. 

\par The calibrations are valid for supergiants with a near-solar metallicity, \Teff\ ranging from 4800 to 6500\,K and \Mv\ from --3 to --6\,mag. The uncertainties due to the effect of luminosity at temperatures above 6200\,K are within 150\,K. The typical mean uncertainty per calibration relation is 60-70\,K (40-45\,K for the most precise ones and 140-160\,K for the least precise ones). Using about 60-70 calibrations improves the intrinsic precision to within 30-50\,K (for spectra with an S/N of 100-150). 

\par Employing this method, we can derive temperatures of highly reddened objects (such as stars towards the Galactic centre). Adopting these calibrations has already enabled us to determine the temperatures of four Cepheids in the Galactic centre discovered by \citet{Matsunaga2011,Matsunaga2013,Matsunaga2015} in order to derive their chemical composition \citep{Kovtyukh2022}. Since many Cepheids have been detected in highly reddened regions, for instance beyond the Galactic center in the far side of the disk \citep[e.g.,][]{Feast2014,Matsunaga2016,Chen2018}, the newly determined LDR will allow us to derive their chemical composition using NIR spectra. To our knowledge, only \citet{Inno2019} tackled this problem so far, determining the metallicity of 5 Cepheids candidates in the inner disk by comparing low-resolution (R$\approx$3\,000) NIR spectra to a pre-computed grid of synthetic spectra.

\par Obtaining spectroscopic time-series for a given star would make it possible to track tiny \Teff\ variations, potentially related to rotational modulation such as those that have already been detected for dwarf stars -- namely, the G8 dwarf $\xi$ Bootis A  \citep{Toner1988}  and the K0 dwarf $\sigma$ Dra  \citep{Gray1992}. 
  This technique is already being used to study spot activity in giants \citep{Berdyugina2005,Fraska2005,Frasca2008}.
In this respect, hemisphere-averaged temperatures of stars with surface inhomogeneities derived 
from NIR lines simultaneously to optical lines can be of great help for starspot modelling.
Indeed, one expects different average temperatures at different wavelengths due to the wavelength 
dependence of the contribution of starspots to the total flux. 

It would be interesting to search simultaneously for systematic variations in spectral line asymmetries in order to better understand the physics of pulsations in Cepheids. As far as Cepheids are concerned, simultaneous time-series spectroscopy in the optical and infrared domain are crucial to refine our understanding of the Cepheids' atmosphere dynamics.

\par In the present paper, the calibration sample is confined to  objects with a near-solar metallicity [Fe/H] to circumvent the issue of the dependence of calibrations on [Fe/H]. Investigating such a dependence of calibrations on [Fe/H] will be the goal of further studies.

 \section{Data Availability}

This research used the facilities of the Italian Center for
Astronomical Archive (IA2) operated by INAF at the Astronomical
Observatory of Trieste, programme \mbox{OPT19A5} (PI: N.Nardetto).
The Subaru/IRCS spectra are available at the SMOKA Science Archive
https://smoka.nao.ac.jp/.
  
\section*{Acknowledgements}
We thank our referee,  Dr. Antonio Frasca, for his important comments, which improved our manuscript.
VK  is grateful to the Vector-Stiftung at Stuttgart, Germany, for support 
within the program "2022--Immediate help for Ukrainian refugee scientists" under 
grant P2022-0064.


\appendix

\section{Calibrations}


\begin{figure*}
\includegraphics[clip=,angle=0,width=0.95\textwidth]{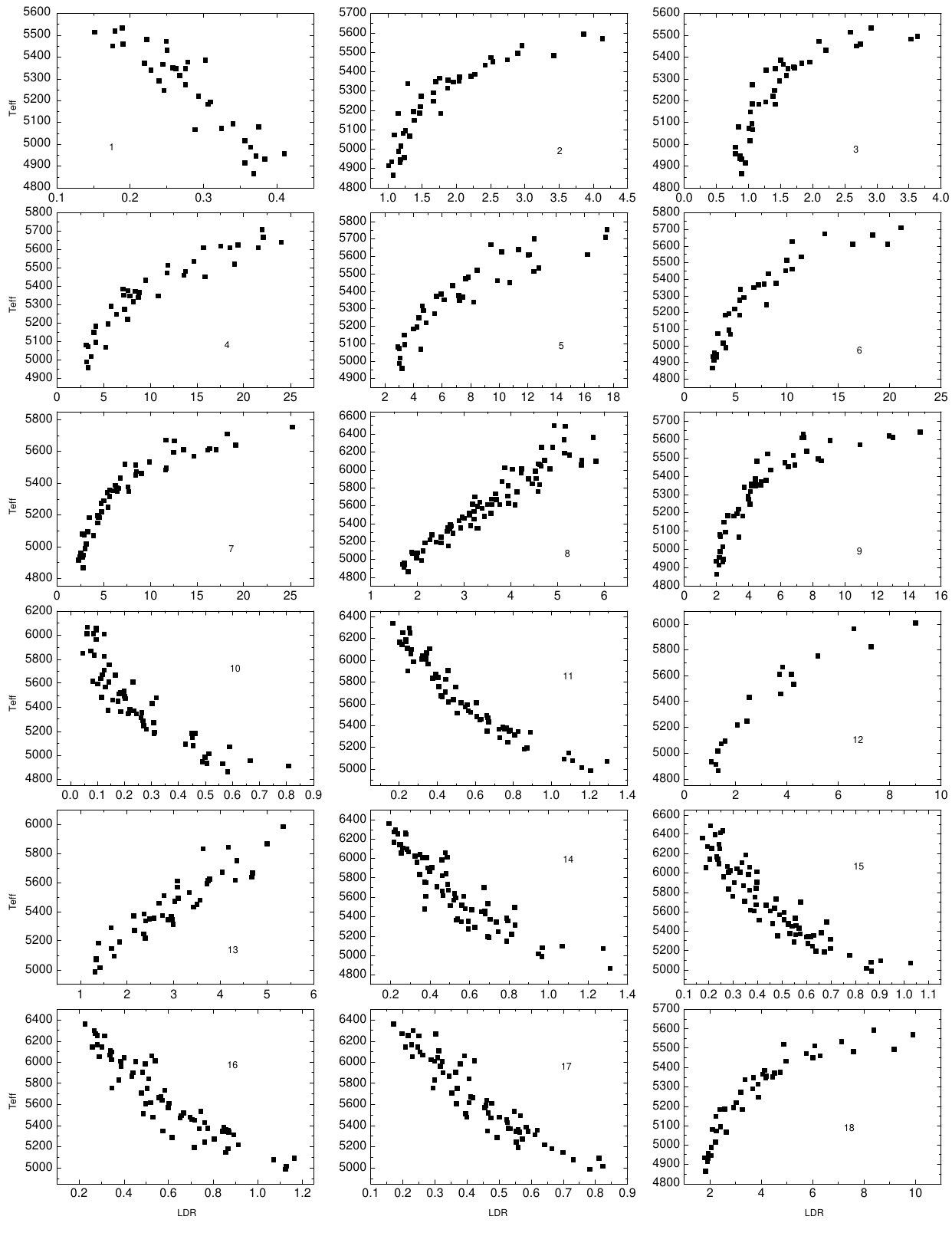}
\caption{Calibration relations 1-18, showing the variation of \Teff\ as a function of the line depth ratio (LDR).}
\label{FigA1}
\end{figure*}

\begin{figure*}
\includegraphics[clip=,angle=0,width=0.95\textwidth]{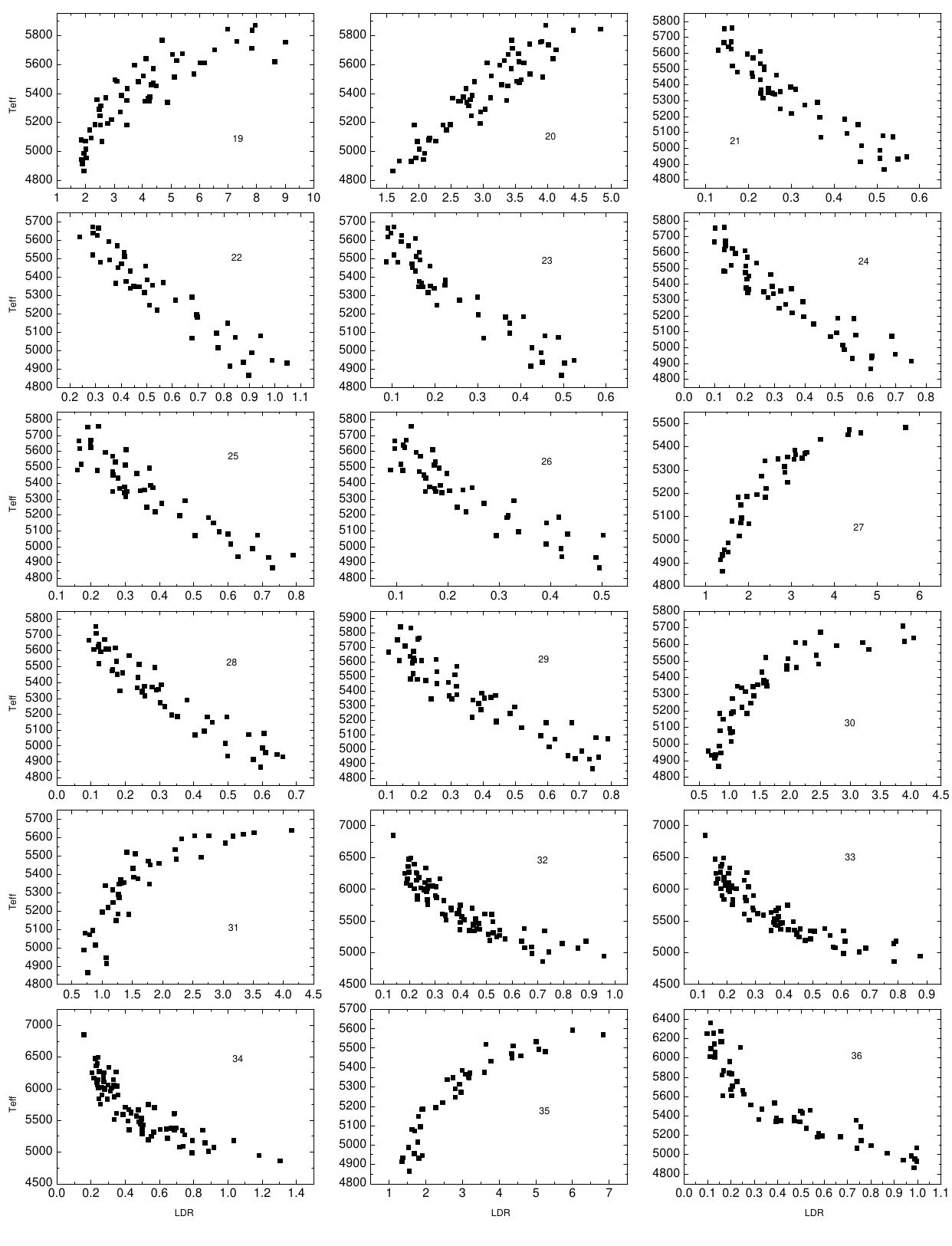}
\caption{Same as in Fig~\ref{FigA1} but for calibration relations 19-36}
\label{FigA2}
\end{figure*}

\begin{figure*}
\includegraphics[clip=,angle=0,width=0.95\textwidth]{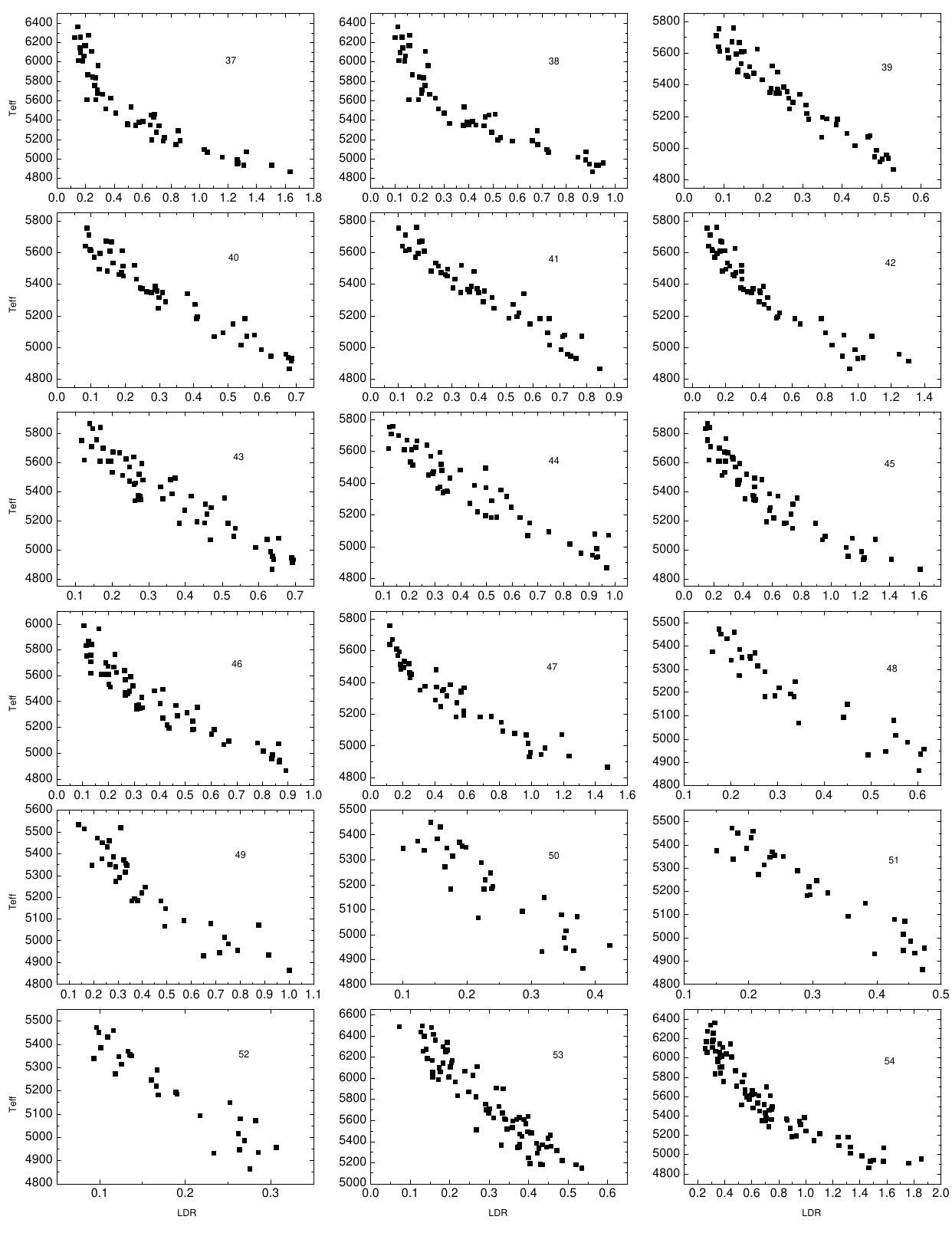}
\caption{Same as in Fig~\ref{FigA1} but for calibration relations 37-54}
\label{FigA3}
\end{figure*}

\begin{figure*}
\includegraphics[clip=,angle=0,width=0.95\textwidth]{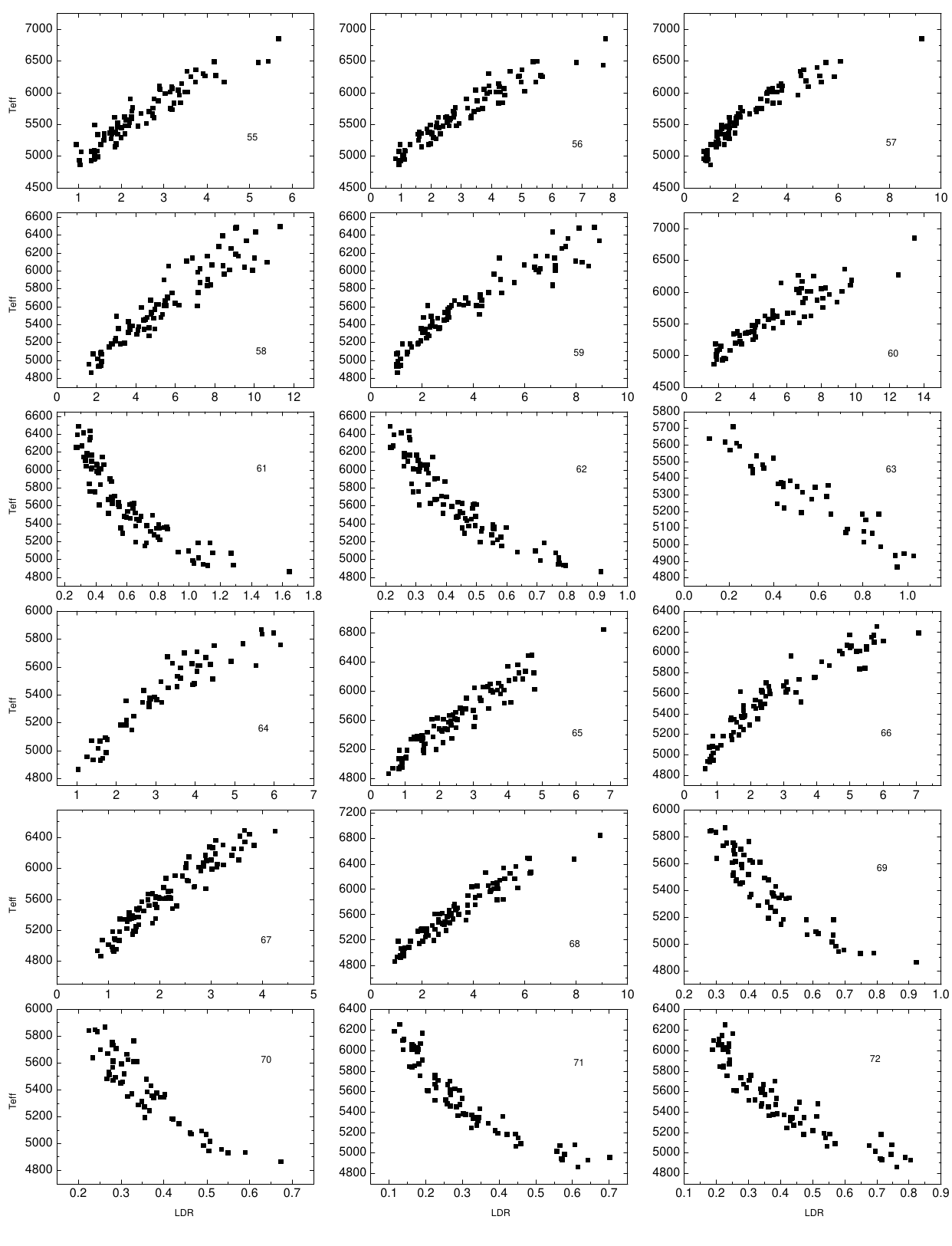}
\caption{Same as in Fig~\ref{FigA1} but for calibration relations 55-72}
\label{FigA4}
\end{figure*}

\begin{figure*}
\includegraphics[clip=,angle=0,width=0.95\textwidth]{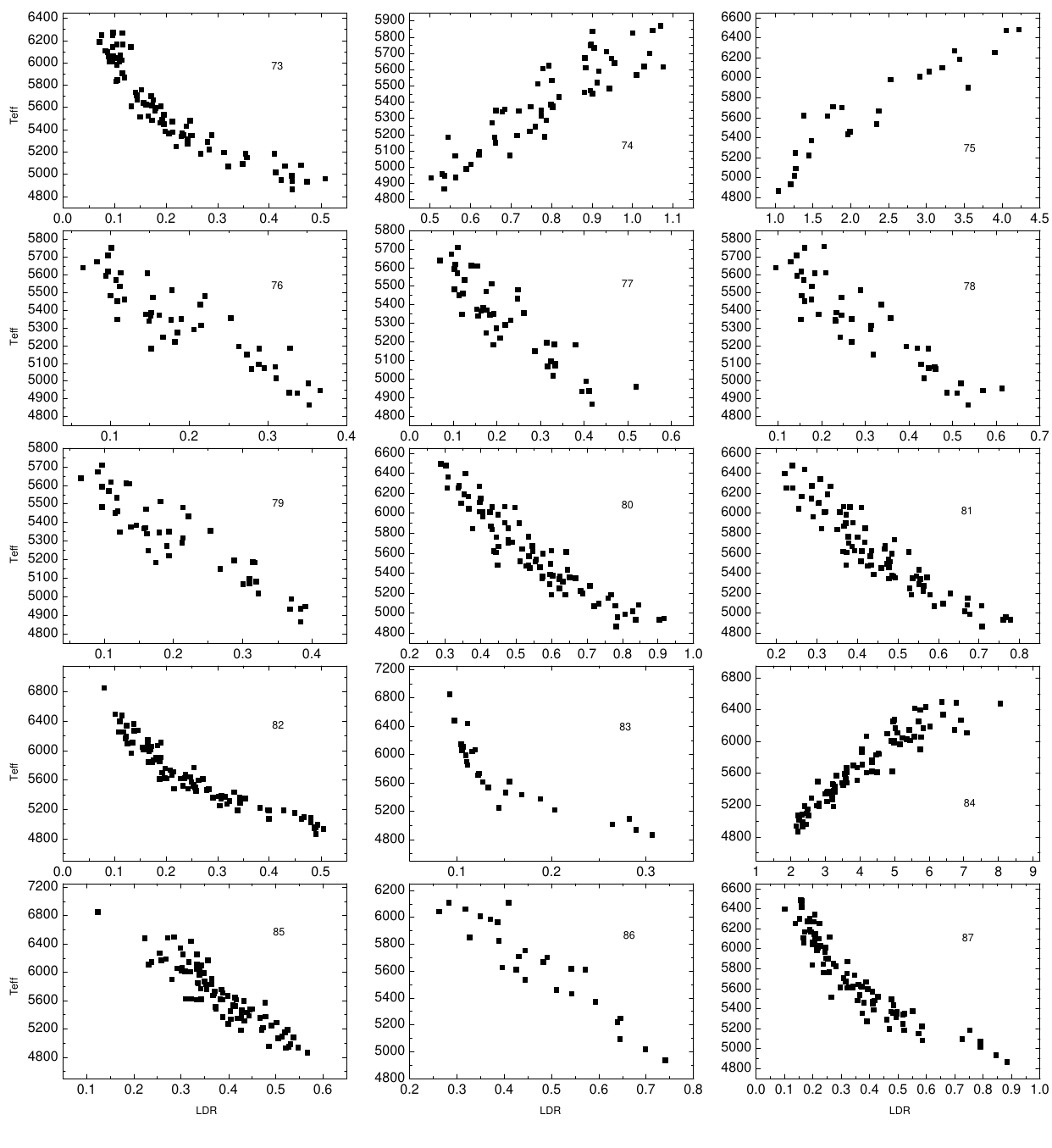}
\caption{Same as in Fig~\ref{FigA1} but for calibration relations 73-87}
\label{FigA5}
\end{figure*}

%
\clearpage


\onecolumn

\begin{longtable}{lcccccrrrrrcr}
\caption{\label{indTeffCal} Individual \Teff\ determinations for calibrating Cepheids. For each individual spectrum, the table provides the observing log, including the pulsation phase of the Cepheid, the temperature and its associated uncertainty, the number of calibrations used for both the calibrating (optical) HARPS-N spectra and the near-infrared GIANO spectra.}\\
\hline\hline 
&&&&&  \multicolumn{4}{c}{GIANO} & \multicolumn{4}{c}{HARPS-N}  \\
  Cep    &   UT Date &   UT     &       JD       & phase & \Teff&$\sigma$&  N    &  $\sigma/\sqrt{N}$      & \Teff&$\sigma$& N  &  $\sigma/\sqrt{N}$     \\
         &           &          &                &       &   (K)  &  (K)&      & (K)    &  (K)    &(K) &     &  (K)    \\

\hline 
  \endfirsthead \caption{continued.}\\ 
  \hline\hline
&&&&&  \multicolumn{4}{c}{GIANO} & \multicolumn{4}{c}{HARPN}  \\
  Cep    &   UT Date &   UT     &       JD       & phase & \Teff&$\sigma$&  N    &  $\sigma/\sqrt{N}$ & \Teff&$\sigma$& N  &       $\sigma/\sqrt{N}$\\
         &           &          &                &       &   (K)  &  (K)&      & (K)    &  (K)    &(K) &     &  (K)    \\
         
   \hline 
\endhead 
$\delta$ Cep& 2019-07-14&03-06-55 &  58678.62980 & .087  &   6219 & 103 & 16   & 25.7   &    6361 &  54&  67 &  6.6    \\
$\delta$ Cep& 2019-07-15&04-30-38 &  58679.68793 & .284  &   5847 &  76 & 35   & 12.9   &    5869 &  65&  79 &  7.3    \\
$\delta$ Cep& 2019-07-16&04-39-15 &  58680.69392 & .471  &   5644 &  88 & 49   & 12.6   &    5618 &  71&  81 &  7.9    \\
$\delta$ Cep& 2019-07-17&02-17-24 &  58681.59541 & .639  &   5537 &  95 & 68   & 11.5   &    5482 &  46&  76 &  5.2    \\
$\delta$ Cep& 2019-07-18&02-38-33 &  58682.61010 & .828  &   5612 & 102 & 43   & 15.6   &    5738 &  73&  79 &  8.2    \\
$\delta$ Cep& 2019-07-19&04-30-31 &  58683.68785 & .029  &   6295 &  87 & 12   & 25.0   &    6486 &  60&  65 &  7.4    \\
$\delta$ Cep& 2019-08-08&03-57-43 &  58703.66508 & .752  &   5513 &  84 & 66   & 10.4   &    5520 &  46&  69 &  5.5    \\
$\delta$ Cep& 2019-08-18&04-54-54 &  58713.70479 & .623  &   5436 & 127 & 59   & 16.6   &    5494 &  61&  77 &  7.0    \\
$\delta$ Cep& 2019-08-21&02-23-57 &  58716.59996 & .162  &   6098 & 137 & 26   & 26.8   &    6145 &  68&  74 &  7.9    \\
$\delta$ Cep& 2019-08-23&00-41-51 &  58718.52906 & .522  &   5548 &  73 & 59   &  9.4   &    5570 &  59&  77 &  6.8    \\
$\delta$ Cep& 2019-08-25&00-31-23 &  58720.52179 & .893  &   6000 & 212 & 17   & 51.5   &    6115 &  91&  74 & 10.6    \\
$\delta$ Cep& 2019-08-26&05-07-52 &  58721.71379 & .115  &   6197 & 120 & 24   & 24.5   &    6273 &  63&  68 &  7.6    \\
$\delta$ Cep& 2019-09-08&00-07-51 &  58734.50545 & .499  &   5559 &  81 & 60   & 10.4   &    5593 &  70&  77 &  7.9    \\
$\delta$ Cep& 2019-09-10&02-48-54 &  58736.61729 & .893  &$\cdots$&$\cdots$&$\cdots$&$\cdots$&6139 & 111&  74 & 12.9    \\
$\delta$ Cep& 2019-09-10&21-51-31 &  58737.41077 & .041  &   6423 & 129 & 15   & 33.2   &    6478 &    &     &         \\
$\delta$ Cep& 2019-09-13&00-10-54 &  58739.50756 & .431  &   5642 &  91 & 51   & 12.7   &    5639 &  82&  80 &  9.2    \\
$\delta$ Cep& 2019-09-21&00-46-38 &  58747.53238 & .927  &   6120 &  97 & 18   & 22.9   &    6339 &  63&  67 &  7.7    \\
         &           &         &              &       &        &     &      &        &         &    &     &         \\
S Sge    & 2019-04-27&05:26:53 &  58600.72700 & .519  &   5474 &  68 & 79   &  7.7   &    5460 & 107&  71 & 12.7    \\
S Sge    & 2019-04-28&05:20:20 &  58601.72245 & .638  &   5314 &  74 & 82   &  8.1   &    5247 & 143&  57 & 18.9    \\
S Sge    & 2019-05-03&05:23:26 &  58606.72460 & .235  &   5942 & 113 & 33   & 19.7   &    5986 & 105&  76 & 12.1    \\
S Sge    & 2019-05-04&05:13:11 &  58607.71748 & .353  &   5704 &  81 & 45   & 12.1   &    5709 & 135&  80 & 15.1    \\
S Sge    & 2019-05-05&05:08:59 &  58608.71457 & .472  &   5529 &  74 & 73   &  8.7   &    5534 & 135&  73 & 15.8    \\
S Sge    & 2019-05-06&05:12:32 &  58609.71703 & .592  &   5355 &  72 & 84   &  7.9   &    5371 & 142&  69 & 17.1    \\
S Sge    & 2019-05-08&05:34:49 &  58611.73251 & .832  &   5779 &  99 & 15   & 25.4   &    5825 & 122&  44 & 18.5    \\
S Sge    & 2019-05-09&05:30:37 &  58612.72959 & .951  &   6221 & 173 & 26   & 33.9   &    6186 & 125&  70 & 15.0    \\
S Sge    & 2019-05-10&05:07:57 &  58613.71385 & .069  &   6130 & 127 & 33   & 22.0   &    6098 & 138&  68 & 16.7    \\
S Sge    & 2019-05-17&03:44:34 &  58620.65594 & .897  &   5898 & 185 & 28   & 35.0   &    5901 & 151&  68 & 18.3    \\
S Sge    & 2019-05-18&03:14:19 &  58621.63494 & .014  &   6204 & 142 & 28   & 26.8   &    6250 & 149&  56 & 19.9    \\
S Sge    & 2019-05-19&03:05:31 &  58622.62883 & .132  &   6013 & 106 & 35   & 17.9   &    6011 & 117&  66 & 14.4    \\
S Sge    & 2019-05-20&03:39:47 &  58623.65262 & .255  &   6018 & 152 & 33   & 26.5   &    6013 & 134&  78 & 15.2    \\
S Sge    & 2019-05-21&03:36:57 &  58624.65065 & .374  &   5666 &  76 & 46   & 11.3   &    5672 & 110&  69 & 13.2    \\
S Sge    & 2019-05-22&03:35:59 &  58625.64998 & .493  &   5492 &  53 & 75   &  6.1   &    5451 & 130&  77 & 14.8    \\
S Sge    & 2019-05-23&03:09:11 &  58626.63137 & .610  &   5290 &  66 & 77   &  7.5   &    5356 & 143&  65 & 17.8    \\
S Sge    & 2019-05-24&03:45:09 &  58627.65635 & .732  &   5354 &  67 & 80   &  7.5   &    5385 & 143&  62 & 18.2    \\
S Sge    & 2019-06-07&02:57:32 &  58641.62328 & .398  &   5578 & 113 & 50   & 16.0   &    5611 & 105&  81 & 11.6    \\
S Sge    & 2019-06-09&02:38:55 &  58643.61035 & .636  &   5322 &  60 & 75   &  6.9   &    5315 & 141&  69 & 17.0    \\
S Sge    & 2019-06-11&02:09:16 &  58645.58976 & .872  &   5823 & 105 & 31   & 18.9   &    5905 & 136&  63 & 17.1    \\
S Sge    & 2019-06-13&01:13:25 &  58647.55098 & .106  &   5964 & 129 & 36   & 21.5   &    6060 & 147&  70 & 17.6    \\
S Sge    & 2019-06-20&00:30:13 &  58654.52098 & .937  &   6026 & 200 & 31   & 35.9   &    6055 & 153&  66 & 18.9    \\
S Sge    & 2019-08-08&03:37:25 &  58703.65098 & .798  &   5492 & 170 & 59   & 22.2   &    5480 & 183&  47 & 26.6    \\
S Sge    & 2019-08-18&22:30:19 &  58714.43771 & .085  &   6001 & 111 & 30   & 20.3   &    6067 & 123&  77 & 14.0    \\
S Sge    & 2019-09-07&22:21:46 &  58734.43178 & .471  &   5496 &  66 & 74   &  7.6   &    5472 &  78&  73 &  9.2    \\
S Sge    & 2019-09-18&23:51:33 &  58745.49413 & .790  &   5555 & 129 & 41   & 20.1   &    5626 & 154&  58 & 20.2    \\
         &           &         &              &       &        &     &      &        &         &    &     &         \\
T Vul    & 2018-08-09&23-17-25 &  58340.47042 & .534  &   5573 & 167 & 39   & 26.7   &    5611 & 146&  52 & 20.2    \\
T Vul    & 2018-08-15&20-53-51 &  58346.37072 & .864  &   6191 & 105 &  9   & 35.0   &    6299 & 157&  37 & 25.8    \\
T Vul    & 2018-08-17&23-15-26 &  58348.46905 & .337  &   5763 & 133 & 32   & 23.5   &    5753 & 124&  66 & 15.3    \\
T Vul    & 2018-08-24&21-01-38 &  58355.37613 & .894  &   6325 & 267 & 11   & 80.6   &    6436 & 125&  41 & 19.4    \\
T Vul    & 2018-08-25&00-46-59 &  58355.53262 & .929  &   6330 & 182 & 13   & 50.5   &    6395 & 146&  53 & 20.1    \\
T Vul    & 2018-08-25&20-49-05 &  58356.36741 & .118  &   6065 & 147 & 22   & 31.3   &    6144 & 137&  70 & 16.3    \\
T Vul    & 2018-08-26&02-18-48 &  58356.59638 & .169  &   5997 & 143 & 32   & 25.2   &    6043 & 137&  54 & 18.7    \\
T Vul    & 2018-08-29&01-41-14 &  58359.57030 & .840  &   6049 & 211 & 20   & 47.2   &    6107 & 143&  53 & 19.7    \\
T Vul    & 2018-09-07&01-42-29 &  58368.57116 & .869  &   6083 & 161 & 22   & 34.3   &    6267 & 155&  58 & 20.3    \\
T Vul    & 2018-09-10&20-25-15 &  58372.35086 & .721  &   5638 & 216 & 22   & 46.0   &    5625 & 124&  55 & 16.7    \\
T Vul    & 2018-09-12&00-55-50 &  58373.53877 & .989  &   6306 & 146 & 13   & 40.4   &    6495 & 177&  48 & 25.5    \\
T Vul    & 2018-09-12&02-57-19 &  58373.62313 & .008  &   6096 & 198 &  5   & 88.6   &    6297 & 138&  55 & 18.6    \\
T Vul    & 2018-10-01&00-07-06 &  58392.50493 & .265  &   5898 & 126 & 10   & 39.9   &    5850 & 113&  79 & 12.8    \\
T Vul    & 2018-12-10&19-15-42 &  58463.30256 & .227  &$\cdots$&$\cdots$&$\cdots$&$\cdots$&5852 & 149&  68 & 18.0    \\
T Vul    & 2018-12-20&19-22-38 &  58473.30738 & .482  &   5599 & 229 & 19   & 52.6   &    5615 & 160&  57 & 21.2    \\
T Vul    & 2019-08-08&03-27-40 &  58703.64421 & .413  &   5681 & 178 & 52   & 24.7   &    5609 & 140&  60 & 18.1    \\
T Vul    & 2019-08-16&20-41-57 &  58712.36246 & .379  &   5712 & 139 & 26   & 27.3   &    5734 & 142&  78 & 16.0    \\
T Vul    & 2019-09-08&01-13-05 &  58734.55075 & .381  &   5634 & 123 & 37   & 20.1   &    5700 & 134&  61 & 17.1    \\
T Vul    & 2019-09-10&20-07-42 &  58737.33868 & .010  &   6362 & 235 & 10   & 74.4   &    6415 & 151&  63 & 19.0    \\
         &           &         &              &       &        &     &      &        &         &    &     &         \\
X Cyg    & 2019-04-27&05-10-08 &  58600.71537 & .360  &   5108 &  40 & 81   &  4.4   &    5094 & 114&  60 & 14.7    \\
X Cyg    & 2019-04-28&05-32-30 &  58601.73090 & .422  &   5047 &  56 & 78   &  6.3   &    5016 &  90&  70 & 10.8    \\
X Cyg    & 2019-05-01&05-37-25 &  58604.73431 & .605  &   4931 &  63 & 18   & 14.8   &    4865 &  90&  52 & 12.5    \\
X Cyg    & 2019-05-03&05-13-56 &  58606.71800 & .726  &   5191 & 143 & 69   & 17.2   &    5183 & 112&  41 & 17.6    \\
X Cyg    & 2019-05-04&04-58-11 &  58607.70707 & .787  &   5416 &  97 & 79   & 10.9   &    5366 &  78&  59 & 10.2    \\
X Cyg    & 2019-05-05&04-54-06 &  58608.70423 & .848  &   5418 &  68 & 81   &  7.6   &    5376 &  77&  67 &  9.4    \\
X Cyg    & 2019-05-06&04-57-28 &  58609.70657 & .909  &   5973 & 111 & 31   & 20.0   &    6008 & 141&  80 & 15.8    \\
X Cyg    & 2019-05-10&05-18-49 &  58613.72140 & .154  &   5400 &  71 & 83   &  7.8   &    5432 & 124&  74 & 14.5    \\
X Cyg    & 2019-05-17&03-30-41 &  58620.64630 & .576  &$\cdots$&$\cdots$&$\cdots$&$\cdots$& 4851 &  76&  57 & 10.0    \\
X Cyg    & 2019-05-18&03-56-20 &  58621.66412 & .638  &$\cdots$&$\cdots$&$\cdots$&$\cdots$& 4913 &  97&  55 & 13.1    \\
X Cyg    & 2019-05-19&03-15-30 &  58622.63576 & .698  &   5126 &  86 & 64   & 10.7   &    5068 & 130&  57 & 17.2    \\
X Cyg    & 2019-05-20&03-29-31 &  58623.64549 & .759  &   5333 &  94 & 73   & 11.0   &    5339 & 104&  53 & 14.3    \\
X Cyg    & 2019-05-21&03-26-24 &  58624.64333 & .820  &   5373 &  89 & 79   & 10.0   &    5346 &  72&  60 &  9.3    \\
X Cyg    & 2019-05-22&03-20-50 &  58625.63946 & .881  &   5638 &  99 & 50   & 14.0   &    5667 &  73&  80 &  8.2    \\
X Cyg    & 2019-05-23&02-59-07 &  58626.62438 & .941  &   6058 & 127 & 24   & 25.8   &    6109 & 137&  72 & 16.1    \\
X Cyg    & 2019-05-24&03-24-59 &  58627.64234 & .003  &   5944 & 132 & 34   & 22.6   &    5963 & 117&  79 & 13.1    \\
X Cyg    & 2019-06-06&04-54-08 &  58640.70425 & .800  &   5368 &  91 & 80   & 10.2   &    5347 &  78&  61 &  9.9    \\
X Cyg    & 2019-08-08&03-46-45 &  58703.65746 & .642  &$\cdots$&$\cdots$&$\cdots$& $\cdots$& 4924 & 111&  55 & 15.0   \\
X Cyg    & 2019-08-22&02-45-33 &  58717.61496 & .494  &   4944 &  46 & 39   &  7.4   &    4935 &  84&  67 & 10.3    \\
X Cyg    & 2019-09-07&23-41-56 &  58734.48745 & .524  &   4933 &  67 & 12   & 19.4   &    4914 &  97&  65 & 12.0    \\
X Cyg    & 2019-09-19&23-41-13 &  58746.48695 & .256  &   5265 &  51 & 73   &  5.9   &    5220 &  72&  72 &  8.5    \\
         &           &         &              &       &        &     &      &        &         &    &     &         \\
SV Vul   & 2019-04-27&04:52:25 &  58600.70306 & .139  &   5967 & 150 & 35   & 25.4   &    5843 & 139&  73 & 16.2    \\
SV Vul   & 2019-04-28&05:10:03 &  58601.71531 & .161  &   5862 & 134 & 36   & 22.4   &    5758 & 137&  81 & 15.2    \\
SV Vul   & 2019-05-01&05:09:53 &  58604.71519 & .228  &   5599 & 137 & 70   & 16.4   &    5513 & 141&  77 & 16.1    \\
SV Vul   & 2019-05-04&05:22:10 &  58607.72372 & .295  &   5431 &  91 & 81   & 10.2   &    5351 & 109&  72 & 12.9    \\
SV Vul   & 2019-05-17&03:53:28 &  58620.66212 & .583  &   5049 &  71 & 73   &  8.4   &    5080 &  84&  61 & 10.7    \\
SV Vul   & 2019-05-21&03:47:34 &  58624.65803 & .672  &   5003 &  89 & 69   & 10.7   &    4987 &  97&  69 & 11.7    \\
SV Vul   & 2019-06-07&03:06:45 &  58641.62968 & .051  &   6133 & 136 & 26   & 26.7   &    6167 & 131&  67 & 16.0    \\  SV Vul   & 2019-06-11&00:58:23 &  58645.54054 & .138  &   5937 & 109 & 36   & 18.1   &    5834 & 125&  71 & 14.8    \\  SV Vul   & 2019-06-20&01:29:58 &  58654.56247 & .339  &   5327 & 113 & 81   & 12.5   &    5290 & 119&  74 & 13.9    \\
SV Vul   & 2019-07-15&03:16:09 &  58679.63621 & .897  &   5323 &  93 & 78   & 10.5   &    5273 &  96&  55 & 12.9    \\
SV Vul   & 2019-07-18&03:27:34 &  58682.64414 & .964  &   6121 & 145 & 25   & 29.0   &    6166 &  88&  52 & 12.2    \\
SV Vul   & 2019-08-08&03:17:46 &  58703.63733 & .432  &   5204 & 105 & 78   & 11.9   &    5185 & 113&  71 & 13.4    \\
SV Vul   & 2019-08-09&21:23:23 &  58705.39123 & .471  &   5153 &  90 & 80   & 10.1   &    5149 & 118&  69 & 14.2    \\
SV Vul   & 2019-08-18&22:21:18 &  58714.43145 & .672  &   4979 & 105 & 46   & 15.4   &    4946 & 109&  57 & 14.5    \\
SV Vul   & 2019-08-21&01:27:25 &  58716.56070 & .720  &   4968 & 102 & 46   & 15.0   &    4932 & 146&  40 & 23.0    \\
SV Vul   & 2019-08-22&20:51:45 &  58718.36927 & .760  &   4937 & 107 & 65   & 13.3   &    4957 & 149&  31 & 26.8    \\
SV Vul   & 2019-08-25&02:36:38 &  58720.60877 & .810  &$\cdots$&$\cdots$&$\cdots$& $\cdots$&4983 & 140&39 & 22.5    \\
SV Vul   & 2019-08-27&20:46:13 &  58723.36542 & .871  &   5137 &  82 & 62   & 10.4   &    5183 & 167&  48 & 24.1    \\
SV Vul   & 2019-08-30&20:20:17 &  58726.34741 & .938  &   5698 & 137 & 14   & 36.6   &    5766 & 138&  52 & 19.2    \\
SV Vul   & 2019-09-03&20:30:20 &  58730.35439 & .027  &   6257 & 158 & 24   & 32.3   &    6253 & 117&  62 & 14.8    \\  
SV Vul   & 2019-09-07&20:35:19 &  58734.35785 & .116  &   6034 & 135 & 26   & 26.4   &    6026 & 160&  42 & 24.7    \\
SV Vul   & 2019-09-19&23:32:17 &  58746.48075 & .386  &   5267 &  89 & 81   &  9.9   &    5194 & 137&  70 & 16.4    \\
         &           &         &              &       &        &     &      &        &         &    &     &         \\
S Vul    & 2019-04-27&04:11:07 &  58600.67438 & .792  &   5282 &  89 & 72   & 10.5   &    5223 & 119&  60 & 15.4    \\
S Vul    & 2019-05-06&05:22:17 &  58609.72380 & .925  &   5856 & 154 & 35   & 26.1   &    5773 & 112&  47 & 16.3    \\
S Vul    & 2019-05-10&04:43:10 &  58613.79664 & .985  &   5990 & 163 & 32   & 28.7   &    5971 & 121&  56 & 16.2    \\
S Vul    & 2019-05-18&03:29:57 &  58621.64579 & .101  &   6069 & 117 & 33   & 20.4   &    5891 & 132&  72 & 15.5    \\
S Vul    & 2019-05-24&03:55:57 &  58627.66385 & .189  &   5900 & 142 & 35   & 24.0   &    5748 & 134&  80 & 15.0    \\
S Vul    & 2019-06-06&05:05:16 &  58640.71199 & .381  &   5478 & 136 & 74   & 15.9   &    5438 & 139&  75 & 16.1    \\
S Vul    & 2019-07-14&02:58:45 &  58678.62413 & .938  &   5822 & 157 & 32   & 27.8   &    5779 & 152&  42 & 23.5    \\
S Vul    & 2019-07-19&04:19:23 &  58683.68012 & .012  &   5965 & 145 & 33   & 25.2   &    5980 & 155&  54 & 21.1    \\
S Vul    & 2019-09-03&21:22:41 &  58730.39075 & .699  &   5201 &  79 & 79   &  8.9   &    5160 & 123&  70 & 14.7    \\
S Vul    & 2019-09-18&23:28:46 &  58745.47831 & .920  &   5548 & 139 & 43   & 21.1   &    5566 & 135&  53 & 18.5    \\
\hline
\end{longtable}



\scriptsize
\begin{longtable}{rclclllrrrrrr}
\caption{\label{tableA1}  LDR-\Teff\ calibration relations. For each relation, the wavelengths of both lines and the corresponding chemical element are provided, together with the analytic function type, the value of the coefficients, the average accuracy of the calibration and the temperature range wherein it can be used. When one of the lines is blended, representing two or more elements, only the elements which predominantly contribute to the blend are indicated.}\\
\hline\hline 
 N &  Lambda1  &  El 1  &  Lambda2 &  El 2     &     Name                  &  \Teff =        & sigma &    \Teff   &     a          &       b      &       c         &      d      \\ 
  & (\AA)  &     &  (\AA) &        &                       &                      & (K) &   (K)   &      &            &             &           \\                   
\hline   
  \endfirsthead \caption{continued.}\\ \hline\hline
 N &  Lambda1  &  El 1  &  Lambda2 &  El 2     &     Name                  &     \Teff =   & sigma &    \Teff   &     a          &       b      &       c         &      d      \\ 
  & (\AA)  &     &  (\AA) &        &                       &                      & (K) &   (K)   &      &            &             &           \\                   
\hline  
\endhead 
 1 &  14968.327& \Fe1      & 15024.992& \Mg1      &   Quadratic Fit           &  a+b$r$+c$r^{2}$        &   81  & 4800--5550  & 	5664.6 &--115.54    &	--4632.1   & $\cdots$  \\
 2 &  14968.327& \Fe1      & 15317.843& \Ti1      &   Modified Exponential    &  a e$^{b/r}$            &   70  & 4950--5550  & 	5831.4 &--0.17068   &  $\cdots$    & $\cdots$  \\
 3 &  15017.700& \Fe1      & 15317.843& \Ti1      &   Modified Hoerl Model    &  ab$^{1/r}r^c$          &   66  & 4900--5550  & 	6022.1 &0.84430     &	--0.032450 & $\cdots$  \\
 4 &  15024.992& \Mg1      & 15221.551& \Ni1      &   Logarithm Fit           &  a+bln($r$)             &   56  & 4950--5700  & 	4662.1 &320.08      &  $\cdots$    & $\cdots$  \\
 5 &  15024.992& \Mg1      & 15328.367& \Sc1      &   Modified Hoerl Model    &  ab$^{1/r}r^c$          &   71  & 5000--5650  & 	5079.3 &0.84156     &	0.044211   & $\cdots$  \\
 6 &  15040.246& \Mg1 +    & 15317.843& \Ti1      &   Hoerl Model             &  a(b$^r$)($r^c$)        &   62  & 4800--5650  & 	4431.1 &0.99541     &	0.11299    & $\cdots$  \\
 7 &  15047.705& \Mg1      & 15317.843& \Ti1      &   Modified Hoerl Model    &  ab$^{1/r}r^c$          &    54 & 4800--5650  & 	5263.3 &0.78937     &	0.030078   & $\cdots$  \\
 8 &  15047.705& \Mg1      & 15387.803& \Fe1      &   Linear Fit              &  a+b$r$                 &   124 & 5000--6250  & 	4367.8 &354.19      &  $\cdots$    & $\cdots$  \\
 9 &  15051.749& \Fe1      & 15317.843& \Ti1      &   Modified Exponential    &  ae$^{b/r}$             &    61 & 4850--5600  & 	5769.8 &--0.32729   &  $\cdots$    & $\cdots$  \\
10 &  15063.513& \Mg1      & 15403.791& \S1       &   Shifted Power Fit       &  a($r$--b)$^c$          &   130 & 4900--5900  & 	4738.9 &--0.050804  &	--0.10471  & $\cdots$  \\
11 &  15077.287& \Fe1      & 15422.276& \S1       &   Quadratic Fit           &  a+b$r$+c$r^2$          &   81  & 5100--6350  & 	6713.5 &--2587.1    &	1006.5     & $\cdots$  \\
12 &  15094.695& \Fe1      & 15317.843& \Ti1      &   Logarithm Fit           &  a+bln($r$)             &   74  & 4800--6000  & 	4843.3 &537.54      &  $\cdots$    & $\cdots$  \\
13 &  15178.422& \V1+\Cr1  & 15210.356& \Sc1 ?    &   Linear Fit              &  a+b$r$                 &   90  & 5000--5950  & 	4816.9 &208.72      &  $\cdots$    & $\cdots$  \\
14 &  15178.422& \V1+\Cr1  & 15403.791& \S1       &   Logarithm Fit           &  a+bln($r$)             &   144 & 4950--6350  & 	5100.0 &--775.93    &  $\cdots$    & $\cdots$  \\
15 &  15178.422& \V1+\Cr1  & 15422.276& \S1       &   Modified Hoerl Model    &  ab$^{1/r}r^c$          &   140 & 5000--6500  & 	4992.1 &0.99911     &	--0.15195  & $\cdots$  \\
16 &  15178.422& \V1+\Cr1  & 15469.816& \S1       &   Logarithm Fit           &  a+bln($r$)             &   126 & 5100--6300  & 	5162.0 &--826.02    &  $\cdots$    & $\cdots$  \\
17 &  15178.422& \V1+\Cr1  & 15478.482& \S1+\Fe1  &   Quadratic Fit           &  a+b$r$+c$r^2$          &   122 & 5100--6300  & 	7008.1 &--3996.6    &	1901.3     & $\cdots$  \\
18 &  15207.526& \Fe1      & 15317.843& \Ti1      &   Modified Exponential    &  ae$^{b/r}$             &   46  & 4850--5550  & 	5730.2 &--0.27878   &  $\cdots$    & $\cdots$  \\
19 &  15207.526& \Fe1      & 15328.367& \Sc1      &   Modified Exponential    &  ae$^{b/r}$             &   105 & 4950--5700  & 	5980.2 &--0.34844   &  $\cdots$    & $\cdots$  \\
20 &  15207.526& \Fe1      & 15490.882& \Fe1 ?    &   Quadratic Fit           &  a+b$r$+c$r^2$          &   100 & 4800--5750  & 	3954.0 &639.38      &	--50.183   & $\cdots$  \\
21 &  15210.356& \Sc1 ?    & 15217.777& \Mn1      &   Quadratic Fit           &  a+b$r$+c$r^2$          &   81  & 4950--5650  & 	6091.5 &--3312.0    &	2270.0     & $\cdots$  \\
22 &  15210.356& \Sc1 ?    & 15224.729& \Fe1      &   Quadratic Fit           &  a+b$r$+c$r^2$          &   75  & 4950--5600  & 	6041.5 &--1678.75   &	569.95     & $\cdots$  \\
23 &  15210.356& \Sc1 ?    & 15376.831& \Si1      &   Quadratic Fit           &  a+b$r$+c$r^2$          &   76  & 4950--5600  & 	5827.4 &--2602.4    &	1730.0     & $\cdots$  \\
24 &  15210.356& \Sc1 ?    & 15400.077& \S1       &   Quadratic Fit           &  a+b$r$+c$r^2$          &   82  & 4950--5600  & 	5884.8 &--2216.1    &	1247.7     & $\cdots$  \\
25 &  15210.356& \Sc1 ?    & 15531.752& \Fe1      &   Quadratic Fit           &  a+b$r$+c$r^2$          &   89  & 4950--5600  & 	5932.8 &--1868.3    &	695.06     & $\cdots$  \\
26 &  15210.356& \Sc1 ?    & 15686.441& \Fe1      &   Quadratic Fit           &  a+b$r$+c$r^2$          &   92  & 5000--5600  & 	5882.1 &--2908.7    &	2148.0     & $\cdots$  \\
27 &  15219.618& \Fe1      & 15317.843& \Ti1      &   Modified Exponential    &  ae$^{(b/r)}$           &   58  & 4900--5450  & 	5721.2 &--0.20768   &  $\cdots$    & $\cdots$  \\
28 &  15221.551& \Ni1      & 15376.831& \Si1      &   Quadratic Fit           &  a+b$r$+c$r^2$          &   74  & 4950--5650  & 	5888.5 &--2320.8    &	1317.5     & $\cdots$  \\
29 &  15221.551& \Ni1      & 15400.077& \S1       &   Quadratic Fit           &  a+b$r$+c$r^2$          &   94  & 4950--5750  & 	5953.6 &--1965.9    &	893.11     & $\cdots$  \\
30 &  15224.729& \Fe1      & 15317.843& \Ti1      &   Modified Exponential    &  ae$^{b/r}$             &   75  & 4950--5600  & 	5836.2 &--0.12437   &  $\cdots$    & $\cdots$  \\
31 &  15239.712& \Fe1      & 15317.843& \Ti1      &   Modified Exponential    &  ae$^{b/r}$             &   96  & 5000--5600  & 	5811.4 &--0.11757   &  $\cdots$    & $\cdots$  \\
32 &  15239.712& \Fe1      & 15403.791& \S1       &   Hoerl Model             &  a(b$^r$)($r^c$)        &   149 & 5050--6900  & 	4446.5 &1.1119      &	--0.20245  & $\cdots$  \\
33 &  15239.712& \Fe1      & 15422.276& \S1       &   Modified Hoerl Model    &  ab$^{1/r}r^c$          &   157 & 5050--6900  & 	4792.2 &1.0139      &	--0.10608  & $\cdots$  \\
34 &  15239.712& \Fe1      & 15469.816& \S1       &   Modified Hoerl Model    &  ab$^{1/r}r^c$          &   180 & 5000--6900  & 	5003.7 &0.99900     &	--0.15005  & $\cdots$  \\
35 &  15244.974& \Fe1      & 15317.843& \Ti1      &   Modified Exponential    &  ae$^{b/r}$             &   60  & 4900--5550  & 	5766.7 &--0.23036   &  $\cdots$    & $\cdots$  \\
36 &  15284.242& \Ti1      & 15422.276& \S1       &   Modified Hoerl Model    &  ab$^{1/r}r^c$          &   132 & 5000--6200  & 	4975.6 &0.99943     &	--0.10023  & $\cdots$  \\
37 &  15284.242& \Ti1      & 15469.816& \S1       &   Modified Hoerl Model    &  ab$^{1/r}r^c$          &   140 & 4850--6250  & 	5105.0 &0.99942     &	--0.10170  & $\cdots$  \\
38 &  15284.242& \Ti1      & 15478.482& \S1 +\Fe1 &   Logarithm Fit           &  a+bln($r$)             &   130 & 4900--6200  & 	4897.0 &--585.43    &  $\cdots$    & $\cdots$  \\
39 &  15317.843& \Ti1      & 15335.383& \Fe1      &   Quadratic Fit           &  a+b$r$+c$r^2$          &   62  & 4800--5700  & 	5877.0 &--2324.2    &	951.69     & $\cdots$  \\
40 &  15317.843& \Ti1      & 15376.831& \Si1      &   Quadratic Fit           &  a+b$r$+c$r^2$          &   59  & 4950--5650  & 	5806.8 &--1748.5    &	677.34     & $\cdots$  \\
41 &  15317.843& \Ti1      & 15400.077& \S1       &   Quadratic Fit           &  a+b$r$+c$r^2$          &   61  & 4950--5650  & 	5820.4 &--1236.2    &	188.60     & $\cdots$  \\
42 &  15317.843& \Ti1      & 15403.791& \S1       &   Shifted Power Fit       &  a($r$--b)$^c$          &   63  & 4950--5700  & 	5077.1 &--0.20476   &	--0.097246 & $\cdots$  \\
43 &  15328.367& \Sc1      & 15376.831& \Si1      &   Quadratic Fit           &  a+b$r$+c$r^2$          &   92  & 4950--5750  & 	5999.2 &--2046.8    &    743.43    & $\cdots$  \\
44 &  15328.367& \Sc1      & 15422.276& \S1       &   Quadratic Fit           &  a+b$r$+c$r^2$          &   79  & 4950--5650  & 	5868.9 &--1414.2    &    487.91    & $\cdots$  \\
45 &  15328.367& \Sc1      & 15469.816& \S1       &   Shifted Power Fit       &  a($r$--b)$^c$          &   75  & 4800--5800  & 	5188.3 &--0.20991   &   --0.10800  & $\cdots$  \\
46 &  15328.367& \Sc1      & 15478.482& \S1 +\Fe1 &   Modified Hoerl Model    &  ab$^{1/r}r^c$          &   122 & 5000--5900  & 	4950.2 &0.99952     &  --0.081113  & $\cdots$  \\
47 &  15363.530& \Ni1 ?    & 15469.816& \S1       &   Hoerl Model             &  a(b$^r$)($r^c$)        &   61  & 4900--5650  & 	5305.5 &0.94907     &  --0.033075  & $\cdots$  \\
48 &  15373.395& \V1 ?     & 15422.276& \S1       &   Power Fit               &  a$r^b$                 &   59  & 4950--5450  & 	4762.4 &--0.076892  &  $\cdots$    & $\cdots$  \\
49 &  15373.395& \V1 ?     & 15469.816& \S1       &   Quadratic Fit           &  a+b$r$+c$r^2$          &   67  & 4950--5500  & 	5753.4 &--1630.6    &	804.19     & $\cdots$  \\
50 &  15373.395& \V1 ?     & 15591.490& \Fe1      &   Linear Fit              &  a+b$r$                 &   75  & 4900--5400  & 	5604.3 &--1695.0    &  $\cdots$    & $\cdots$  \\
51 &  15373.395& \V1 ?     & 15604.221& \Fe1      &   Linear Fit              &  a+b$r$                 &   58  & 4900--5450  & 	5712.6 &--1637.6    &  $\cdots$    & $\cdots$  \\
52 &  15373.395& \V1 ?     & 15748.988& Mg 1      &   Linear Fit              &  a+b$r$                 &   66  & 4900--5400  & 	5653.2 &--2445.6    &  $\cdots$    & $\cdots$  \\
53 &  15381.960& \Fe1      & 15422.276& \S1       &   Quadratic Fit           &  a+b$r$+c$r^2$          &   135 & 5100--6600  & 	6901.6 &--4616.3    &	2415.2     & $\cdots$  \\
54 &  15387.803& \Fe1      & 15403.791& \S1       &   3rd degree Polynom. Fit &  a+b$r$+c$r^2$+d$r^3$   &   111 & 4900--6350  & 	6954.8 &--3223.0    &	1889.5     & --404.99  \\
55 &  15400.077& \S1       & 15490.882& \Fe1   ?  &   Quadratic Fit           &  a+b$r$+c$r^2$          &   145 & 4700--6900  & 	4316.8 &654.34      &	--42.104   & $\cdots$  \\
56 &  15403.791& \S1       & 15490.882& \Fe1   ?  &   Quadratic Fit           &  a+b$r$+c$r^2$          &   122 & 4700--6900  & 	4602.7 &434.24      &	--22.149   & $\cdots$  \\
57 &  15403.791& \S1       & 15665.241& \Fe1      &   Modified Hoerl Model    &  ab$^{1/r}r^c$          &   127 & 4700--6900  & 	5087.6 &1.00095     &	0.12909    & $\cdots$  \\
58 &  15422.276& \S1       & 15485.454& \Fe1      &   Quadratic Fit           &  a+b$r$+c$r^2$          &   143 & 4800--6300  & 	4534.9 &243.68      &	--6.8790   & $\cdots$  \\
59 &  15422.276& \S1       & 15514.279& \Fe1      &   Shifted Power Fit       &  a($r$--b)$^c$          &   114 & 4900--6200  & 	4555.3 &--0.86381   &	0.14224    & $\cdots$  \\
60 &  15422.276& \S1       & 15674.653& \Si1      &   Linear Fit              &  a+b$r$                 &   170 & 5000--6900  & 	4801.6 &147.88      &  $\cdots$    & $\cdots$  \\
61 &  15459.343& \Y1+\V1 ? & 15469.816& \S1       &   Hoerl Model             &  a(b$^r$)($r^c$)        &   157 & 5000--6600  & 	4699.9 &1.0884      &	--0.22551  & $\cdots$  \\
62 &  15459.343& \Y1+\V1 ? & 15478.482& \S1+\Fe1  &   Hoerl Model             &  a(b$^r$)($r^c$)        &   137 & 4900--6500  & 	4695.0 &1.0074      &	--0.20295  & $\cdots$  \\
63 &  15462.287& \Co1 ?    & 15519.361& \Fe1      &   Linear Fit              &  a+b$r$                 &   74  & 4900--5700  & 	5748.9 &--824.48    &  $\cdots$    & $\cdots$  \\
64 &  15469.816& \S1       & 15485.454& \Fe1      &   Power Fit               &  a$r$$^b$               &   81  & 4900--5600  & 	4787.8 &0.11016    &  $\cdots$    & $\cdots$  \\
65 &  15469.816& \S1       & 15490.882& \Fe1   ?  &   Shifted Power Fit       &  a($r$--b)$^c$          &   128 & 4700--6900  & 	2834.5 &--3.8666    &	0.37210    & $\cdots$  \\
66 &  15469.816& \S1       & 15514.279& \Fe1      &   Hoerl Model             &  a(b$^r$)($r^c$)        &   95  & 4900--6200  & 	5067.7 &1.0068      &	0.083165   & $\cdots$  \\
67 &  15469.816& \S1       & 15611.045& \V1       &   Quadratic Fit           &  a+b$r$+c$r^2$          &   125 & 4800--6500  & 	4312.3 &771.88      &	--60.147   & $\cdots$  \\
68 &  15478.482& \S1       & 15490.882& \Fe1   ?  &   Quadratic Fit           &  a+b$r$+c$r^2$          &   121 & 4700--6900  & 	4581.2 &379.54      &	--15.872   & $\cdots$  \\
69 &  15490.882& \Fe1      & 15604.221& \Fe1      &   Quadratic Fit           &  a+b$r$+c$r^2$          &   100 & 4900--5800  & 	6858.0 &--4248.8    &	2260.6     & $\cdots$  \\
70 &  15490.882& \Fe1      & 15621.654& \Fe1      &   Quadratic Fit           &  a+b$r$+c$r^2$          &   107 & 4900--5750  & 	6878.4 &--5478.4    &	3604.0     & $\cdots$  \\
71 &  15514.279& \Fe1      & 15591.490& \Fe1      &   Hoerl Model             &  a(b$^r$)($r^c$)        &   98  & 4900--6150  & 	4394.5 &1.0669      &	--0.16310  & $\cdots$  \\
72 &  15514.279& \Fe1      & 15604.221& \Fe1      &   Hoerl Model             &  a(b$^r$)($r^c$)        &   116 & 5000--6100  & 	4371.5 &1.1056      &	--0.19259  & $\cdots$  \\
73 &  15514.279& \Fe1      & 15748.988& \Mg1      &   Hoerl Model             &  a(b$^r$)($r^c$)        &   135 & 5000--6300  & 	3963.4 &1.2149      &	--0.17771  & $\cdots$  \\
74 &  15519.096& \Fe1      & 15665.241& \Fe1      &   Logarithm Fit           &  a+bln($r$)             &   117 & 4900--5750  & 	5702.7 &1195.2      &  $\cdots$    & $\cdots$  \\
75 &  15652.871& \Fe1      & 15680.069& \Cr1      &   Logarithm Fit           &  a+bln($r$)             &   162 & 5000--6300  & 	4914.1 &1020.1      &  $\cdots$    & $\cdots$  \\
76 &  15658.545& \Ti1      & 15661.898& \Ti1+\Fe1 &   Linear Fit              &  a+b$r$                 &   103 & 4950--5500  & 	5791.4 &--2352.5    &  $\cdots$    & $\cdots$  \\
77 &  15658.545& \Ti1      & 15906.044& \Fe1      &   Linear Fit              &  a+b$r$                 &   104 & 4950--5500  & 	5727.0 &--1825.1    &  $\cdots$    & $\cdots$  \\
78 &  15658.545& \Ti1      & 15912.591& \Fe1      &   Linear Fit              &  a+b$r$                 &   108 & 4950--5500  & 	5807.8 &--1596.5    &  $\cdots$    & $\cdots$  \\
79 &  15658.545& \Ti1      & 15920.637& \Fe1      &   Linear Fit              &  a+b$r$                 &   92  & 4950--5500  & 	5765.5 &--2154.6    &  $\cdots$    & $\cdots$  \\
80 &  15665.240& \Fe1      & 15677.519& \Fe1      &   Power Fit               &  a$r$$^b$               &   137 & 4900--6650  & 	4769.5 &--0.24874   &  $\cdots$    & $\cdots$  \\
81 &  15665.240& \Fe1      & 15686.441& \Fe1      &   Quadratic Fit           &  a+b$r$+c$r^2$          &   144 & 4900--6450  & 	7531.9 &--5704.6    &	2994.0     & $\cdots$  \\
82 &  15665.240& \Fe1      & 15748.988& \Mg1      &   Shifted Power Fit       &  a($r$--b)$^c$          &   103 & 5000--6900  & 	4460.3 &0.017860    &	--0.15095  & $\cdots$  \\
83 &  15673.538& \Mn1 +    & 15748.988& \Mg1      &   Modified Hoerl Model    &  ab$^{1/r}r^c$          &   164 & 4900--6900  & 	5152.3 &1.0760      &	0.22764    & $\cdots$  \\
84 &  15748.988& \Mg1      & 15941.848& \Fe1      &   Modified Hoerl Model    &  ab$^{1/r}r^c$          &   137 & 4900--6900  & 	6192.5 &0.53702     &	0.056421   & $\cdots$  \\
85 &  15941.848& \Fe1      & 15980.726& \Fe1      &   Linear Fit              &  a+b$r$                 &   179 & 4700--6900  &   7359.4 &--4381.17   &  $\cdots$    & $\cdots$  \\
86 &  16584.447& \V1       & 16632.019& \Mg1 +\Fe1&   Linear Fit              &  a+b$r$                 &   124 & 4900--6100  & 	6793.3 &--2441.3    &  $\cdots$    & $\cdots$  \\
87 &  16645.874& \Fe1      & 16890.414& \Fe1      &   Modified Hoerl Model    &  ab$^{1/r}r^c$          &   125 & 5000--6500  & 	4791.8 &0.99914     &	--0.15115  & $\cdots$  \\
\hline
\end{longtable}                                                          

\bsp	

\label{lastpage}

\begin{thebibliography}{99}

\bibitem[\protect\citeauthoryear{Af{\c{s}}ar et al.}{2023}]{Afsar2023} 
        Af{\c{s}}ar M.,  Bozkurt  Z., Topcu G. B., {\"O}zdemir S.,  Sneden  Chr.,  Mace  G. N., Jaffe D. T., L{\'o}pez-Valdivia R., 2023, ApJ, 949, 86A 


\bibitem[\protect\citeauthoryear{Alonso et al.}{1996}]{Alonso1996} 
        Alonso  A.,  Arribas  S.,  Martinez-Roger  C., 1996, A\&A, 313, 873 

\bibitem[\protect\citeauthoryear{Anderson}{2016}]{Anderson2016} 
        Anderson  R. I., 2016, MNRAS 463, 1707 

\bibitem[\protect\citeauthoryear{Andrievsky et al.}{2002a}]{Andrievsky2002a}
Andrievsky S.~M., Kovtyukh V.~V., Luck R.~E., Lepin\'e J.~R.~D., Bersier D., Maciel W.~J., Barbuy B., 
Klochkova V.G., Panchuk V.~E., Karpischek R.~U., 2002a, A\&A, 381, 32

\bibitem[\protect\citeauthoryear{Andrievsky et al.}{2002b}]{Andrievsky2002b}
Andrievsky S.~M., Bersier D., Kovtyukh V.~V., Luck R.~E., Maciel W.~J., Lepnin\'e J.~R.~D., 
Beletsky Yu.~V., 2002b, A\&A, 384, 140

\bibitem[\protect\citeauthoryear{Andrievsky et al.}{2002c}]{Andrievsky2002c}
Andrievsky S.~M., Kovtyukh V.~V., Luck R.~E.,  Le\'epine J.~R.~D., Maciel W.~J., Beletsky Yu.~V., 
2002c, A\&A, 392, 491

\bibitem[\protect\citeauthoryear{Andrievsky et al.}{2004}]{Andrievsky2004}
Andrievsky S.~M., Luck R.~E., Martin P., Le\'epine J.~R.~D., 2004, A\&A, 413, 159

\bibitem[\protect\citeauthoryear{Andrievsky et al.}{2005}]{Andrievsky2005}
    Andrievsky S. M., Luck R. E., Kovtyukh V. V., 2005, AJ, 130, 1880

\bibitem[\protect\citeauthoryear{Berdyugina et al.}{2005}]{Berdyugina2005}
    Berdyugina S. V., 2005, Living Rev. Sol. Phys., 2, 8

\bibitem[\protect\citeauthoryear{Bessell et al.}{1998}]{Bessell1998} 
   Bessell M. S., Castelli F., Plez B., 1998, A\&A, 333, 231

\bibitem[\protect\citeauthoryear{Biazzo et al.}{2004}]{Biazzo2004}
   Biazzo K., Catalano S., Frasca A., Marilli E., 2004, Mem. Soc. Astron. Ital. Supplementi, 5, 109

\bibitem[\protect\citeauthoryear{Biazzo  et al.}{2006}]{Biazzo2006}
    Biazzo K. , Frasca A., Catalano S., Marilli E., 2006, preprint 
    (arXiv astro-ph/0610584)

\bibitem[\protect\citeauthoryear{Biazzo et al.}{2007}]{Biazzo2007}
   Biazzo K., Frasca A., Catalano S., Marilli E., 2007, Astron. Nachr., 328, 938

\bibitem[\protect\citeauthoryear{Blackwell \& Shallis}{1977}]{Blackwell1977}
   Blackwell D. E., Shallis M. J., 1977, MNRAS, 180, 177

\bibitem[\protect\citeauthoryear{Caccin et al.}{2002}]{Caccin2002}
    Caccin B., Penza V., Gomez M. T., 2002, A\&A, 386, 286

\bibitem[\protect\citeauthoryear{Catalano et al.}{2002}]{Catalano2002}
    Catalano S., Biazzo K., Frasca A., Marilli E., 2002, A\&A, 394, 1009

\bibitem[\protect\citeauthoryear{Cayrel \& Cayrel}{1963}]{Cayrel1963}
    Cayrel G., Cayrel R., 1963, ApJ, 137, 431

\bibitem[\protect\citeauthoryear{Chen et al.}{2018}]{Chen2018}
   Chen X., Wang S., Deng L., de Grijs R., Yang M., 2018, ApJS, 237, 28

\bibitem[\protect\citeauthoryear{Cosentino et al.}{2012}]{Cosentino2012}
    Cosentino R. et al., 2012, in McLean I. S., Ramsay S. K., Takami H.eds, Proc. SPIE Conf. Ser. Vol 8446, Ground-based and Airborne Instrumentation for Astronomy IV. SPIE, Bellingham. p. 84461V

\bibitem[\protect\citeauthoryear{da Silva et al.}{2016}]{daSilva2016}
    da Silva R. et al., 2016, A\&A, 586, A125

\bibitem[\protect\citeauthoryear{da Silva et al.}{2022}]{daSilva2022}
     da Silva R. et al., 2022, A\&A, 661, A104

\bibitem[\protect\citeauthoryear{Davis \& Webb}{1974}]{Davis1974}
    Davis J., Webb R. J., 1974, MNRAS, 168, 163

\bibitem[\protect\citeauthoryear{Feast et al.}{2014}]{Feast2014}
   Feast M. W., Menzies J. W., Matsunaga N., Whitelock P. A., 2014, Nature, 509, 342

\bibitem[\protect\citeauthoryear{Fraska et al.}{2005}]{Fraska2005}
   Frasca A., Biazzo K., Catalano S., Marilli E., Messina S., Rodon{\`o} M., 2005, A\&A, 432, 647

\bibitem[\protect\citeauthoryear{Frasca et al.}{2008}]{Frasca2008}
   Frasca A., Biazzo K., Ta{\c{s}} G., Evren S., Lanzafame A. C., 2008, A\&A, 479, 557

\bibitem[\protect\citeauthoryear{Fukue et al.}{2015}]{Fukue2015}
    Fukue K. et al., 2015, ApJ, 812, 64

\bibitem[\protect\citeauthoryear{Gehren}{1981}]{Gehren1981}
        Gehren T., 1981, A\&A 100, 97

\bibitem[\protect\citeauthoryear{Genovali et al.}{2013}]{Genovali2013} 
   Genovali K., et al, 2013, 554, 132

\bibitem[\protect\citeauthoryear{Genovali et al.}{2014}]{Genovali2014} 
   Genovali K., et al, 2014, 566, 37

\bibitem[\protect\citeauthoryear{Genovali et al.}{2015}]{Genovali2015} 
   Genovali K., et al, 2015, 580, 17

\bibitem[\protect\citeauthoryear{Gray}{1989}]{Gray1989} 
        Gray D. F., 1989,ApJ 347, 1021

\bibitem[\protect\citeauthoryear{Gray \& Johanson}{1991}]{Gray1991}
       Gray  D. F., Johanson H. L., 1991, PASP 103 439

\bibitem[\protect\citeauthoryear{Gray et al.}{1992}]{Gray1992}
        Gray, D. F., Baliunas, S. L., Lockwood   G.~W., Skiff B. A., 1992, ApJ 400, 681

\bibitem[\protect\citeauthoryear{Gray}{1994}]{Gray1994}
       Gray  D. F., 1994, PASP 106, 1248

\bibitem[\protect\citeauthoryear{Gray \& Brown}{2001}]{Gray2001}
       Gray  D. F., Brown  K., 1994, PASP 113, 723

\bibitem[\protect\citeauthoryear{Gray}{2005}]{Gray2005}
       Gray  D. F.,  The Observation and Analysis of Stellar Photospheres,
        2005, 3rd edn., Cambridge Univ. Press, Cambridge, UK

\bibitem[\protect\citeauthoryear{Hanke et al.}{2018}]{Hanke2018}
   Hanke M., Hansen C. J., Koch A., Grebel E. K., 2018, A\&A, 619, A134

\bibitem[\protect\citeauthoryear{Inno et al.}{2019}]{Inno2019}
     Inno L. et al., 2019, MNRAS 482, 83

\bibitem[\protect\citeauthoryear{Jian et al.}{2019}]{Jian2019}
    Jian M., Matsunaga N., Fukue K., 2019, MNRAS, 485, 1310

\bibitem[\protect\citeauthoryear{Jian et al.}{2020}]{Jian2020}
       Jian M. et al., 2020, MNRAS 494, 1724

\bibitem[\protect\citeauthoryear{Kobayashi et al.}{2000}]{Kobayashi2000}
   Kobayashi N. et al., 2000, in Iye M., Moorwood A. F.eds, Proc. SPIE Conf. Ser. Vol. 4008, Optical and IR Telescope Instrumentation and Detectors. SPIE, Bellingham. p. 1056

\bibitem[\protect\citeauthoryear{Kovtyukh}{2007}]{Kovtyukh2007}
 Kovtyukh V.~V., 2007, MNRAS, 378, 617

\bibitem[\protect\citeauthoryear{Kovtyukh \& Andrievsky}{1999}]{Kovtyukh1999}
   Kovtyukh V. V. , Andrievsky S. M., 1999, A\&A, 351, 597

\bibitem[\protect\citeauthoryear{Kovtyukh et al.}{2019}]{Kovtyukh19}
   Kovtyukh V. et al., 2019, MNRAS, 488, 3211

\bibitem[\protect\citeauthoryear{Kovtyukh \& Gorlova}{2000}]{Kovtyukh2000}
     Kovtyukh V. V., Gorlova N. I., 2000, A\&A, 358, 587

\bibitem[\protect\citeauthoryear{Kovtyukh et al.}{2003}]{Kovtyukh2003}
     Kovtyukh V. V., Soubiran C., Belik S. I., Gorlova N. I., 2003, A\&A, 411, 559

\bibitem[\protect\citeauthoryear{Kovtyukh et al.}{2005a}]{Kovtyukh2005a} 
    Kovtyukh V.~V., Andrievsky S.~M., Belik S.~I., Luck R.~E., 
      AJ, 129, 433
 
\bibitem[\protect\citeauthoryear{Kovtyukh, Wallerstein, \& Andrievsky}{2005b}]{Kovtyukh2005b} 
  Kovtyukh V.~V., Wallerstein G., Andrievsky S.~M., 2005, PASP, 117, 1173

\bibitem[\protect\citeauthoryear{Kovtyukh, Wallerstein, \& Andrievsky}{2005c}]{Kovtyukh2005c} 
  Kovtyukh V.~V., Wallerstein G., Andrievsky S.~M., 2005, PASP, 117, 1182


\bibitem[\protect\citeauthoryear{Kovtyukh  et al.}{2006}]{Kovtyukh2006}
    Kovtyukh V. V., Soubiran C., Bienaym{\'e} O., Mishenina T. V., Belik S. I., 2006, MNRAS, 371, 879

\bibitem[\protect\citeauthoryear{Kovtyukh et al.}{2016}]{Kovtyukh2016}
  Kovtyukh V. et al., 2016, MNRAS, 460, 2077

\bibitem[\protect\citeauthoryear{Kovtyukh et al.}{2018a}]{Kovtyukh2018a}
    Kovtyukh V. et al., 2018a, PASP, 130, 54201

\bibitem[\protect\citeauthoryear{Kovtyukh et al.}{2018b}]{Kovtyukh2018b}
    Kovtyukh V., Yegorova I., Andrievsky S., Korotin S., Saviane I., Lemasle B., Chekhonadskikh F., Belik S.,    2018b, MNRAS, 477, 2276

\bibitem[\protect\citeauthoryear{Kovtyukh et al.}{2019}]{Kovtyukh2019}
    Kovtyukh V. et al., 2019, MNRAS, 488, 3211

\bibitem[\protect\citeauthoryear{Kovtyukh et al.}{2022}]{Kovtyukh2022}
     Kovtyukh V. V., Korotin S. A., Andrievsky S. M., Matsunaga N., Fukue K., 2022, MNRAS, 516, 4269

\bibitem[\protect\citeauthoryear{Lemasle et al.}{2007}]{Lemasle2007}
     Lemasle B., Fran{\c c}ois P., Bono G., Mottini M., Primas F., Romaniello M., 2007, A\&A, 467, 283

\bibitem[\protect\citeauthoryear{Lemasle et al.}{2008}]{Lemasle2008}
   Lemasle B., Fran{\c c}ois P., Piersimoni A., Pedicelli S., Bono G., Laney C. D., Primas F., Romaniello M., 2008, A\&A , 490, 613

\bibitem[\protect\citeauthoryear{Lemasle et al.}{2013}]{Lemasle2013}
   Lemasle B. et al., 2013, A\&A, 558, A31

\bibitem[\protect\citeauthoryear{Lemasle et al.}{2015}]{Lemasle2015}
    Lemasle B. et al., 2015, A\&A , 579, A47

\bibitem[\protect\citeauthoryear{Lemasle  et al.}{2017}]{Lemasle2017}
    Lemasle B. et al., 2017, A\&A, 608, A85

\bibitem[\protect\citeauthoryear{Lemasle  et al.}{2018}]{Lemasle2018}
    Lemasle B. et al., 2018, A\&A, 618, A160

\bibitem[\protect\citeauthoryear{Lemasle et al.}{2020}]{Lemasle2020}
   Lemasle B. , Hanke M., Storm J., Bono G., Grebel E. K., 2020, A\&A, 641, A71

\bibitem[\protect\citeauthoryear{Luck}{2018}]{Luck2018b}
   Luck R. E., 2018, AJ, 156, 171

\bibitem[\protect\citeauthoryear{Luck \& Lambert}{2011}]{Luck2011b}
   Luck R. E., Lambert D. L., 2011, AJ, 142, 136

\bibitem[\protect\citeauthoryear{Luck \& Andrievsky}{2004}]{Luck2004}
     Luck R. E., Andrievsky S. M., 2004, AJ, 128, 343

\bibitem[\protect\citeauthoryear{Luck  et al.}{2003}]{Luck2003}
   Luck R. E. , Gieren W. P., Andrievsky S. M., Kovtyukh V. V., Fouqu{\'e} P., Pont F., Kienzle F., 2003, A\&A, 401, 939

\bibitem[\protect\citeauthoryear{Luck et al.}{2006}]{Luck2006}
   Luck R. E. , Kovtyukh V. V., Andrievsky S. M., 2006, AJ, 132, 902

\bibitem[\protect\citeauthoryear{Luck et al.}{2008}]{Luck2008}
   Luck R. E., Andrievsky S. M., Fokin A., Kovtyukh V. V., 2008, AJ, 136, 98

\bibitem[\protect\citeauthoryear{Luck et al.}{2011}]{Luck2011a}
     Luck R. E., Andrievsky S. M., Kovtyukh V. V., Gieren W., Graczyk D., 2011, AJ, 142, 51

\bibitem[\protect\citeauthoryear{Martin et al.}{2015}]{Martin2015}
   Martin R. P., Andrievsky S. M., Kovtyukh V . V ., Korotin S. A., Yegorova I. A., Saviane I., 2015, MNRAS, 449, 4071

\bibitem[\protect\citeauthoryear{Matsunaga et al.}{2011}]{Matsunaga2011} 
Matsunaga N.,  Kawadu T., Nishiyama Sh., Nagayama T., Kobayashi N., Tamura M., 
Bono G., Feast M.~W., Nagata T., 2011, Nature, 477, Iss. 7363, 188  

\bibitem[\protect\citeauthoryear{Matsunaga et al.}{2013}]{Matsunaga2013} 
        Matsunaga  N.,  Feast  M.  W., Kawadu T., Nishiyama Sh., Nagayama T., Nagata T., Tamura M., Bono G.,     Kobayashi N., 2013, MNRAS 429, 385

\bibitem[\protect\citeauthoryear{Matsunaga et al.}{2015}]{Matsunaga2015} 
Matsunaga N., Fukue K., Yamamoto R., Kobayashi N., Inno L., Genovali K., Bono G., 
et al., 2015, ApJ, 799, 46

\bibitem[\protect\citeauthoryear{Matsunaga et al.}{2016}]{Matsunaga2016} 
Matsunaga N., Feast M.~W., Bono G., Kobayashi N., Inno L., Nagayama T., Nishiyama S., 
et al., 2016, MNRAS, 462, 414.

\bibitem[\protect\citeauthoryear{Matsunaga}{2017}]{Matsunaga2017} 
        Matsunaga N., 2017, in European Physical Journal Web of Conferences, 152, p.01007 
         preprint ( arXiv:1705.02547 )

\bibitem[\protect\citeauthoryear{Matsunaga et al.}{2021}]{Matsunaga2021}
        Matsunaga N., Jian M., Taniguchi D., Elgueta S S., 2021, MNRAS 506, 1031

\bibitem[\protect\citeauthoryear{Nardetto et al.}{2006}]{Nardetto2006}
   Nardetto N., Mourard D., Kervella P., Mathias P., M{\'e}rand A., Bersier D., 2006, A\&A, 453, 309

\bibitem[\protect\citeauthoryear{Nardetto et al.}{2007}]{Nardetto2007}
     Nardetto N., Mourard D., Mathias P., Fokin A., Gillet D., 2007, A\&A, 471, 661

\bibitem[\protect\citeauthoryear{Nardetto et al.}{2008a}]{Nardetto2008a}
     Nardetto N., Stoekl A., Bersier D., Barnes T. G., 2008a, A\&A, 489, 1255

\bibitem[\protect\citeauthoryear{Nardetto et al.}{2008b}]{Nardetto2008b}
     Nardetto N., Groh J. H., Kraus S., Millour F., Gillet D., 2008b, A\&A, 489, 1263

\bibitem[\protect\citeauthoryear{Nardetto et al.}{2018}]{Nardetto2018}
    Nardetto N. et al., 2018, A\&A, 616, A92

\bibitem[\protect\citeauthoryear{Nardetto et al.}{2023}]{Nardetto2023}
    Nardetto N. et al., 2023, A\&A , 671, A14

\bibitem[\protect\citeauthoryear{Ness et al.}{2015}]{Ness2015}
     Ness M., Hogg D. W., Rix H. W., Ho A. Y. Q., Zasowski G., 2015, ApJ, 808, 16

\bibitem[\protect\citeauthoryear{Origlia et al.}{2014}]{Origlia2014}
     Origlia L. et al., 2014, in Ramsay S. K., McLean I. S., Takami H. eds, Proc. SPIE Conf. Ser. Vol. 9147, Ground-based and Airborne Instrumentation for Astronomy V. SPIE, Bellingham. p. 91471E

\bibitem[\protect\citeauthoryear{Pedicelli et al.}{2010}]{Pedicelli2010}
    Pedicelli S. et al., 2010, A\&A, 518, A11

\bibitem[\protect\citeauthoryear{Proxauf et al.}{2018}]{Proxauf2018}
       Proxauf B. et al., 2018, A\&A, 616, A82

\bibitem[\protect\citeauthoryear{Recio-Blanco et al.}{2006}]{Recio-Blanco2006}
       Recio-Blanco A., Bijaoui A., Laverny P., 2006, MNRAS, 370, 141

\bibitem[\protect\citeauthoryear{Romaniello et al.}{2008}]{Romaniello2008}
    Romaniello M. et al., 2008, A\&A, 488, 731

\bibitem[\protect\citeauthoryear{Romaniello et al.}{2022}]{Romaniello2022}
   Romaniello M. et al., 2022, A\&A, 658, A29

\bibitem[\protect\citeauthoryear{Strassmeier \& Schordan}{2000}]{Strassmeier2000}
            Strassmeier K. G., Schordan P., 2000, Astron. Nachr., 321, 277

\bibitem[\protect\citeauthoryear{Taniguchi et al.}{2018}]{Taniguchi2018}
    Taniguchi D. et al., 2018, MNRAS, 473, 4993

\bibitem[\protect\citeauthoryear{Taniguchi et al.}{2021}]{Taniguchi2021}
    Taniguchi D. et al., 2021, MNRAS, 502, 4210

\bibitem[\protect\citeauthoryear{Toner \& Gray}{1988}]{Toner1988} 
   Toner C.~G., Gray D. F., 1988, ApJ 334, 1008

\bibitem[\protect\citeauthoryear{van Hoof \& Struve}{1953}]{VanHoof1953} 
        van Hoof A. Struve O., 1953, PASP 65, 158 

\bibitem[\protect\citeauthoryear{Vasilyev et al.}{2017}]{Vasilyev2017}
   Vasilyev V.  Ludwig  H.-G.  Freytag B. Lemasle B.  Marconi M., 2017, A\&A, 606, 140

\bibitem[\protect\citeauthoryear{Vasilyev et al.}{2018}]{Vasilyev2018}
       Vasilyev V., Ludwig H. -G.,  Freytag B.,  Lemasle B., Marconi M., A\&A 611, 19

\bibitem[\protect\citeauthoryear{Vasilyev et al.}{2019}]{Vasilyev2019}
    Vasilyev V.,  Amarsi A.~M.,  Ludwig, H. -G., Lemasle B., A\&A 624, 85



\end{thebibliography}
\end{document}